\documentclass[aps,pre,twocolumn,superscriptaddress]{revtex4-1}

\usepackage{euscript,amsmath,amssymb,amsfonts,graphicx,bm,amsthm,mathtools}

  \usepackage{paralist}
  \usepackage{epstopdf}
  \usepackage{graphics} 
 \usepackage[latin1]{inputenc}
\usepackage[caption=false]{subfig}
\usepackage{soul}

\usepackage{latexsym,amssymb,enumerate,amsmath,amsthm} 
  \usepackage{graphicx} 
  \usepackage{tikz}
  \usepackage[hidelinks]{hyperref}
\usepackage{latexsym,amssymb,enumerate,amsmath} 
\usepackage{pgfplots}
\usepackage{soul,xcolor}

\newcommand{\grad}{\nabla}
\newcommand{\dd}{\textup{d}}
\def\eps{\varepsilon}
\def\E{\mathbb{E}}
\def\P{\mathbb{P}}
\def\R{\mathbb{R}}

\theoremstyle{plain}
\theoremstyle{remark}

\theoremstyle{definition}

\begin{document}


\title[]{Stochastic search with space-dependent diffusivity}


\author{Hwai-Ray Tung}
\affiliation{University of Utah, Department of Mathematics, Salt Lake City, UT 84112 USA}
\author{Sean D. Lawley}
\email[]{lawley@math.utah.edu}
\affiliation{University of Utah, Department of Mathematics, Salt Lake City, UT 84112 USA}


\date{\today}

\begin{abstract}
The canonical model of stochastic search tracks a randomly diffusing ``searcher'' until it finds a ``target.'' Owing to its many applications across science and engineering, this perennially popular problem has been thoroughly investigated in a variety of models. However, aside from some exactly solvable one-dimensional examples, very little is known if the searcher diffusivity varies in space. For such space-dependent or ``heterogeneous'' diffusion, one must specify the interpretation of the multiplicative noise, which is termed the It\^{o}-Stratonovich dilemma. In this paper, we investigate how stochastic search with space-dependent diffusivity depends on this interpretation. We obtain general formulas for the probability distribution and all the moments of the stochastic search time and the so-called splitting probabilities assuming that the targets are small or weakly reactive. These asymptotic results are valid for general space-dependent diffusivities in general domains in any space dimension with targets of general shape which may be in the interior or on the boundary of the domain. We illustrate our theory with stochastic simulations. Our analysis predicts that stochastic search can depend strongly and counterintuitively on the multiplicative noise interpretation.
\end{abstract}


\maketitle

\section{\label{intro}Introduction}

Stochastic search pervades natural and engineered systems \cite{grebenkov2024target}. As examples, the ``searcher'' and ``target'' could be a ligand and a receptor \cite{shoup82}, a sperm and an egg \cite{meerson2015}, or a predator and a prey \cite{kurella2015}. The search speed is typically quantified by the first passage time (FPT), which is the stochastic time that it takes the searcher to ``find'' the target. 

The canonical model of stochastic search tracks a randomly diffusing searcher in a bounded spatial domain. Aside from some exactly solvable special cases in simple geometries, most analytical results on the FPT were obtained in the regime that the targets are small compared to the length scale of the domain, which is termed the narrow escape or narrow capture problem \cite{benichou2008narrow, grebenkov2016, agranov2018narrow, holcman2014, ward10, ward10b}. The narrow escape problem models a variety of biophysical processes, especially in cell biology \cite{holcman2014}. While most studies assume that the searcher moves by simple diffusion, the narrow escape framework has been extended to more complicated searcher dynamics, including stochastically gated targets \cite{PB3}, fluctuating diffusivity \cite{lawley2019dtmfpt}, and stochastic resetting \cite{bressloff2020search}.

One natural extension of the canonical diffusive search model is a diffusion coefficient $D(x)$ which depends on the spatial location $x$ of the searcher. Important theoretical advances related to ergodicity \cite{cherstvy2013anomalous, cherstvy2014particle, massignan2014nonergodic, xu2020heterogeneous, wang2020anomalous, wang2021time} and FPTs in radially symmetric domains \cite{godec2015optimization, vaccario2015, godec2016first} for space-dependent diffusivities have been made. Nevertheless, stochastic search with a space-dependent diffusivity remains poorly understood, despite its ubiquity for protein diffusion inside of biological cells \cite{english2011single, kuhn2011protein, cutler2013multi, manzo2015review, wu2018, pacheco2024langevin}. 

Spatially dependent diffusion is sometimes termed ``heterogeneous diffusion'' or ``nonlinear Brownian motion'' \cite{klimontovich1994nonlinear}. In this model, the position $X(t)$ of the searcher at time $t$ evolves according to the following stochastic differential equation,
\begin{align}\label{eq:sde0}
    X(t+\dd t)
    =X(t)+\sqrt{2D(X^{*})}\,\dd W(t),
\end{align}
where $\dd W$ is a Brownian increment and the space-dependent diffusivity $D(x)$ is evaluated at the following weighted average of the current searcher position $X(t)$ and its position at the next infinitesimal time $X(t+\dd t)$,
\begin{align}\label{eq:weighted0}
    X^{*}
    =(1-\alpha)X(t)
    +\alpha X(t+\dd t),
\end{align}
for a given ``interpretation parameter'' $\alpha\in[0,1]$. The probability density for the searcher position,
\begin{align}\label{eq:density0}
    p(x,t)
    =\frac{\P(X(t)=x)}{\dd x},
\end{align}
evolves according to the following $\alpha$-dependent forward Fokker-Planck equation \cite{bressloff2017temporal},
\begin{align}\label{eq:fpe0}
    \partial_t p
    =\grad\cdot[(D(x))^\alpha\grad[(D(x))^{1-\alpha}p]].
\end{align}

Due to the ``jaggedness'' of Brownian paths (precisely, their nonzero quadratic variation \cite{oksendal2003}), the choice of $\alpha$ can strongly affect the searcher dynamics.
The three most common choices are $\alpha=0$ \cite{ito1944stochastic}, $\alpha=1/2$ \cite{stratonovich1966new}, and $\alpha=1$ \cite{hanggi1982stochastic}, which are respectively termed the It\^{o}, Stratonovich, and kinetic interpretations. The kinetic interpretation is also called isothermal, H{\"a}nggi-Klimontovich, or anti-It\^{o}. Despite the It\^{o} versus Stratonovich controversy of the 1970s and 1980s \cite{van1981ito} (and more recent discussions prompted by advances in experimental techniques \cite{volpe2010influence, mannella2011comment, volpe2011volpe, bhattacharyay2025brownian}), it is now generally agreed that there is no universally ``correct'' choice of $\alpha\in[0,1]$ \cite{mannella2012ito, sokolov2010ito}. That is, the interpretation parameter ${\alpha}\in[0,1]$ is part of the model and must be chosen on physical grounds \cite{van1981ito, mannella2012ito}.

\begin{figure}
\centering
\includegraphics[width=1\linewidth]{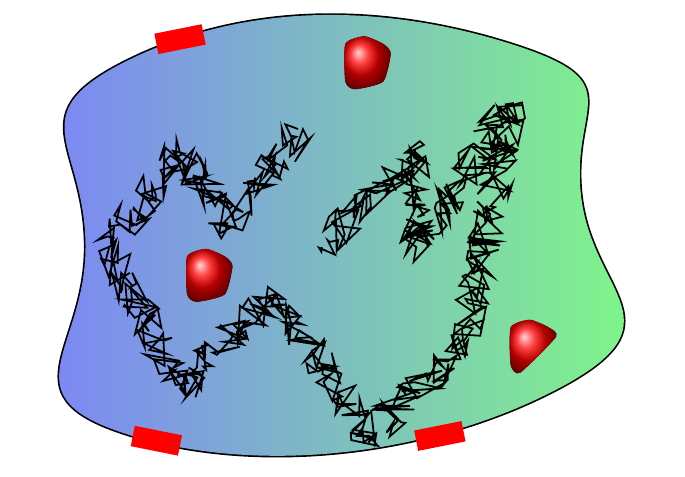}
\caption{A searcher diffuses (black path) with a space-dependent diffusivity (blue-green gradient) inside a general $d$-dimensional domain with small targets (red regions) on its boundary and its interior.}
\label{fig:schem}
\end{figure}

In this paper, we study how stochastic search with a space-dependent diffusivity depends on the interpretation parameter $\alpha\in[0,1]$. We find that search statistics can depend strongly and counterintuitively on $\alpha$. We consider search with a general space-dependent diffusivity $D(x)$ in a general $d$-dimensional bounded domain $\Omega\subset\R^d$ and derive asymptotic results assuming that the $N\ge1$ targets are small and/or weakly reactive (see Figure~\ref{fig:schem} for an illustration). In particular, we analytically derive all the moments of the FPT and its limiting probability distribution, the mean residence time in any subdomain $\Omega'\subset\Omega$, and the probability that the searcher finds a specified target, which is called the ``splitting probability'' \cite{redner2001}. We consider targets which are either perfectly absorbing (which means that the searcher ``finds'' or is ``absorbed'' at a target immediately upon first contact) or imperfectly absorbing (which means that the searcher must accumulate enough time near the target to be absorbed). The targets can be located on the boundary of the confining domain $\Omega$ or in the interior of $\Omega$, and the targets have general shapes. We reported some of these results in Ref.~\cite{tung2025escape} (specifically, the results on the mean FPT and splitting probability for a three-dimensional domain with disk-shaped targets on the boundary). Our analysis relies on a spectral solution of the forward equation \eqref{eq:fpe0} (and the backward equation), eigenvalue asymptotics, and strong localized perturbation theory \cite{ward1993summing, ward1993strong}.

The rest of the paper is organized as follows. We formulate the model in section~\ref{sec:model} and analyze it in sections~\ref{sec:math}-\ref{sec:higherorder}. 
We compare our analytical results to stochastic simulations in section~\ref{sec:sims}. We compare the It\^{o}, Stratonovich, and kinetic interpretations in 
section~\ref{sec:implications}. We conclude by discussing relations to prior work.

\section{\label{sec:model}Model formulation}

\subsection{Stochastic differential equation}
Suppose $X=\{X(t)\}_{t\ge0}$ is the $d$-dimensional heterogeneous diffusion process described by \eqref{eq:sde0}-\eqref{eq:fpe0}, where $D:\R^d\to\R$ is a given space-dependent diffusivity. Assume $X$ diffuses in a bounded $d$-dimensional domain $\Omega\subset\R^d$ with reflecting boundary conditions. Suppose the boundary $\partial\Omega$ contains $N\ge1$ distinguished regions $\partial\Omega_1,\dots,\partial\Omega_N$ which we refer to as the targets. Note that these targets may be ``boundary targets'' in that they are on the outer boundary of $\Omega$, or they may be ``interior targets'' or ``holes'' in the interior of the domain. See Figure~\ref{fig:schem} for an illustration.

If the targets are ``perfect'' or ``perfectly absorbing,'' then the FPT to absorption at a target is the following random time,
\begin{align}\label{eq:tau}
    \tau
    :=\inf\{t\ge0:X(t)\in\cup_{n=1}^N\partial\Omega_n\}.
\end{align}
If the targets are ``imperfect'' or ``partially absorbing,'' then the searcher $X$ may not be absorbed upon its first encounter with a target. Precisely, the FPT to absorption for imperfect targets is the following random time,
\begin{align}\label{eq:taukappa}
    \tau
    :=\inf\{t\ge0:\ell_n(t)>E_n/\kappa_n\text{ for some }n\in\{1,\dots,N\}\},
\end{align}
where $\ell_n(t)$ is the local time \cite{grebenkov2006} of $X(t)$ on $\partial\Omega_n$, $E_1,\dots,E_N$ are independent unit rate exponential random variables, and $\kappa_n>0$ is the reactivity of the $n$th target.

Finally, the ``residence time'' or ``occupation time'' of $X$ in a subset $\Omega'\subseteq\Omega$ is 
\begin{align}\label{eq:residencedefn}
    \tau_{\Omega'}
    :=\int_0^\tau 1_{X(t)\in\Omega'}\,\dd t,
\end{align}
where $1_{X(t)\in\Omega'}=1$ if $X(t)\in\Omega'$ and $1_{X(t)\in\Omega'}=0$ otherwise. 
In words, $\tau_{\Omega'}$ is the total time that $X$ spends in $\Omega'$ before absorption. Naturally, the residence time in the entire domain is the FPT, $\tau_{\Omega}=\tau$.

\subsection{Forward equation}
The probability density function of $X(t)$ in \eqref{eq:density0} evolves according to the forward Fokker-Planck (or forward Kolmogorov) equation in \eqref{eq:fpe0}, which can be written as
\begin{align*}
    \partial_t p
    =\mathcal{L}^{}p,\quad x\in\Omega,
\end{align*}
where $\mathcal{L}^{}$ is the forward operator,
\begin{align}\label{eq:forwardL}
    \mathcal{L}^{}p
    =\grad\cdot[D^\alpha\grad[D^{1-\alpha}p]].
\end{align}
Furthermore, $p$ satisfies reflecting boundary conditions away from the targets,
\begin{align}\label{eq:preflect}
    -D^\alpha\grad[D^{1-\alpha}p]\cdot\mathbf{n}(x)
    =0,\quad x\in\partial\Omega\backslash\{\cup_{n=1}^N\partial\Omega_n\},
\end{align}
where $\mathbf{n}(x)$ is the unit normal at $x\in\partial\Omega$. If the targets are perfect, then $p$ satisfies absorbing boundary conditions at each target,
\begin{align}\label{eq:pabsorb}
    p
    =0,\quad x\in\cup_{n=1}^N\partial\Omega_n.
\end{align}
If the targets are imperfect, then $p$ satisfies partially absorbing boundary conditions at each target,
\begin{align}\label{eq:ppartial}
    -D^\alpha\grad[D^{1-\alpha}p]\cdot\mathbf{n}(x)
    =\kappa_i p ,\quad x\in\partial\Omega_{i}.
\end{align}
If $X(0)=x_0\in\Omega$, then $p$ has the Dirac initial condition,
\begin{align*}
    p(x,0)
    =\delta(x-x_0),\quad x\in\Omega.
\end{align*}

\subsection{Backward equation}
The heterogeneous diffusion process $X$ can also be described by its survival probability,
\begin{align}\label{eq:surv}
    S(x,t)
    =\P_x(\tau>t),
\end{align}
where $\tau$ is the absorption time of $X$ in \eqref{eq:tau} or \eqref{eq:taukappa} and $\P_x$ denotes probability conditioned on the initial location $X(0)=x$. The survival probability evolves according to the backward Fokker-Planck equation,
\begin{align*}
    \partial_t S
    =\mathcal{L}^{*}S,\quad x\in\Omega,
\end{align*}
where $\mathcal{L}^{*}$ is the backward operator, which is the adjoint of the forward operator $\mathcal{L}^{}$ in \eqref{eq:forwardL},
\begin{align}\label{eq:backwardL}
    \mathcal{L}^{*}S
    =D^{1-\alpha}\grad\cdot[D^\alpha\grad S],
\end{align}
and $S$ satisfies adjoint boundary conditions to \eqref{eq:pabsorb}-\eqref{eq:ppartial}. Specifically, $S$ satisfies reflecting boundary conditions away from targets,
\begin{align}\label{eq:Sreflect}
    \partial_{\mathbf{n}}S
    =0,\quad x\in\partial\Omega\backslash\{\cup_{n=1}^N\partial\Omega_n\},
\end{align}
where $\partial_{\mathbf{n}}S=\grad S\cdot\mathbf{n}$ denotes the normal derivative. At the targets, $S$ satisfies either absorbing boundary conditions for perfect targets,
\begin{align}\label{eq:Sabsorb}
    S
    =0,\quad x\in\cup_{n=1}^N\partial\Omega_n,
\end{align}
or partially absorbing boundary conditions if the targets are imperfect,
\begin{align}\label{eq:Spartial}
    -D\partial_{\mathbf{n}}S
    =\kappa_i S ,\quad x\in\partial\Omega_{i}.
\end{align}
Since the process cannot be absorbed immediately, the survival probability satisfies the unit initial condition,
\begin{align*}
    S=1,\quad t=0,\,x\in\Omega.
\end{align*}

\subsection{Spectral expansions}\label{sec:spectral}

For $\beta\in[0,1]$, define the weighted inner product,
\begin{align*}
(f,g)_{\beta}
:=\int_{\Omega}f(x)g(x)(D(x))^{\beta}\,\dd x,
\end{align*} 
on the associated weighted space of square integrable functions, $L_\beta^2(\Omega)=\{f:(f,f)_\beta<\infty\}$. It is straightforward to check that the forward operator $\mathcal{L}^{}$ in \eqref{eq:forwardL} is formally self-adjoint on $L_{1-\alpha}^2(\Omega)$ with the boundary conditions for $p$ in \eqref{eq:preflect}-\eqref{eq:ppartial}. Similarly, the backward operator $\mathcal{L}^{*}$ in \eqref{eq:backwardL} is formally self-adjoint on $L_{\alpha-1}^2(\Omega)$ with the boundary conditions for $S$ in \eqref{eq:Sreflect}-\eqref{eq:Spartial}. We therefore let
\begin{align*}
    0<\lambda_0<\lambda_1<\cdots
\end{align*}
denote the positive eigenvalues of $-\mathcal{L}^{}$ with orthogonal eigenfunctions ${u}^{}_n$ satisfying
\begin{align}\label{eq:feval}
    \lambda_n{u}^{}_n
    =-\mathcal{L}^{}{u}^{}_n,\quad x\in\Omega,
\end{align}
with the same boundary conditions as $p$ in \eqref{eq:preflect}-\eqref{eq:ppartial}. We normalize these forward eigenfunctions so that
\begin{align*}
    (u_n^{},u_m^{})_{1-\alpha}
    =\delta_{nm}\int_{\Omega} u_n^{} u_m^{} D^{1-\alpha}\,\dd x
    =\delta_{nm}\int_{\Omega} D^{\alpha-1}\,\dd x,
\end{align*}
where $\delta_{nm}$ denotes the Kronecker delta. 
It is straightforward to check that the backward operator $\mathcal{L}^{*}$ in \eqref{eq:backwardL} has the same eigenvalues,
\begin{align}\label{eq:beval}
    \lambda_n{u}_n^{*}
    =-\mathcal{L}^{*}{u}_n^{*},\quad x\in\Omega,
\end{align}
where the orthogonal eigenfunctions of $\mathcal{L}^{*}$ satisfy the same boundary conditions as $S$ in \eqref{eq:Sabsorb}-\eqref{eq:Spartial} and are related to the forward eigenfunctions via
\begin{align}\label{eq:dual}
    {u}_n^{*}
    =D^{1-\alpha}{u}^{}_n,\quad n\ge0,
\end{align}
and satisfy
\begin{align*}
    ({u}_n^{*},{u}_m^{*})_{\alpha-1}
    =\delta_{nm}\int_\Omega D^{\alpha-1}\,\dd x.
\end{align*}
We thus formally expand $p(x,t)$ and $S(x,t)$,
\begin{align}
    p(x,t)
    &=\frac{D(x_0)^{1-\alpha}}{\int_\Omega D^{\alpha-1}\,\dd y}\sum_{n\ge0}e^{-\lambda_n t}u_n^{}(x_0) {u}_n^{}(x),\label{eq:pexpand}\\
    S(x,t)
    &=\frac{1}{\int_\Omega D^{\alpha-1}\,\dd y}\sum_{n\ge0}e^{-\lambda_n t}({u}_n^{*},1)_{\alpha-1} {u}_n^{*}(x).\label{eq:Sexpand}
\end{align}

\subsection{Slow escape statistics}

We are interested in parameter regimes in which the principal eigenvalue $\lambda_0$ vanishes, which corresponds to a slow escape. In particular, if the targets are all small and/or weakly reactive, then the principal eigenvalue $\lambda_0$ vanishes and the principal eigenfunctions approach the corresponding Neumann eigenfunctions for the unperturbed (i.e.\ targetless) problem,
\begin{align}\label{eq:slow}
    u_0^{}\to D^{\alpha-1}
    \quad\text{and}\quad
    {u}_0^{*}\to1\quad\text{as }\lambda_0\to0.
\end{align}
Integrating the forward eigenvalue equation in \eqref{eq:feval} for $n=0$, using the dual eigenfunction relationship in \eqref{eq:dual}, and applying the divergence theorem and the boundary conditions in \eqref{eq:Sreflect}-\eqref{eq:Sabsorb} implies that
\begin{align*}
    -\lambda_0\int_\Omega{{u}^{}_0}\,\dd x
    &=\int_\Omega \grad\cdot[D^\alpha\grad{u}_0^{*}]\,\dd x
    =\sum_{j=1}^N\int_{\partial\Omega_j}D^\alpha\partial_{\mathbf{n}} {u}_0^{*}\,\dd x.
\end{align*}
Therefore, in the slow escape regime, \eqref{eq:slow} implies
\begin{align}\label{eq:lambdaasymptotic}
    \lambda_0
    \sim\Big(\int_\Omega D^{\alpha-1}\,\dd x\Big)^{-1}\sum_{j=1}^N\int_{\partial\Omega_j}D^\alpha\partial_{\mathbf{n}} {u}_0^{*}\,\dd x.
\end{align}
We now use \eqref{eq:lambdaasymptotic} to show how knowledge of the flux of ${u}_0^{*}$ into the targets yields information about statistics of the diffusion process $X$.

The $m$th moment of the FPT $\tau$ in \eqref{eq:tau}-\eqref{eq:taukappa} is obtained by integrating the survival probability $S$ in \eqref{eq:surv}
\begin{align*}
    \E_x[\tau^m]
    &=m\int_0^\infty t^{m-1}S(x,t)\,\dd t\\
    &=\frac{m!}{\int_\Omega D^{\alpha-1}\,\dd y}\sum_{n\ge0}\frac{1}{\lambda_n^m}({u}_n^{*},1)_{\alpha-1} {u}_n^{*}(x),
\end{align*}
where the second equality uses the spectral expansion in \eqref{eq:Sexpand} and $\E_x$ denotes expectation conditioned that $X(0)=x$. 
In the slow escape regime of vanishing $\lambda_0$, the principal eigenfunction asymptotics in \eqref{eq:slow} imply that the moments of the FPT $\tau$ become identical to an exponential random variable with rate $\lambda_0$,
\begin{align}\label{eq:moments}
    \E_x[\tau^m]
    &\sim\frac{m!}{\lambda_0^m}\quad\text{as }\lambda_0\to0.
\end{align}
In fact, \eqref{eq:slow} implies that $\lambda_0\tau$ converges to a unit rate exponential random variable in the slow escape regime of $\lambda_0\to0$,
\begin{align}\label{eq:convdist}
\begin{split}
    &\P_x(\lambda_0 \tau>s)
    =S(x,s/\lambda_0)\\
    &=\frac{1}{\int_\Omega D^{\alpha-1}\,\dd x}\sum_{n\ge0}e^{-\frac{\lambda_n }{\lambda_0}s}({u}_n^{*},1)_{\alpha-1} {u}_n^{*}(x)
    \to e^{-s}.
\end{split}    
\end{align}

The mean of the residence time $\tau_{\Omega'}$ in \eqref{eq:residencedefn} in a subset $\Omega'\subseteq\Omega$ is obtained by integrating the density $p$ in \eqref{eq:density0},
\begin{align*}
    \E_{x_0}[\tau_{\Omega'}]
    &=\int_0^\infty\int_{\Omega'}p(x,t\,|\,x_0)\,\dd x\,\dd t\\
    &=\frac{D(x_0)^{1-\alpha}}{\int_\Omega D^{\alpha-1}\,\dd x}\sum_{n\ge0}\frac{1}{\lambda_n}u_n^{}(x_0)\int_{\Omega'}u_n^{}\,\dd x,
\end{align*}
where the second equality uses the spectral expansion in \eqref{eq:pexpand}. 
In the slow escape regime, \eqref{eq:slow} implies that 
\begin{align}\label{eq:residence}
    \E_{x_0}[\tau_{\Omega'}]
    \sim\frac{1}{\lambda_0}\frac{\int_{\Omega'} D^{\alpha-1}\,\dd x}{\int_\Omega D^{\alpha-1}\,\dd x}\quad\text{as }\lambda_0\to0.
\end{align}

The probability that $X$ finds the $j$th target before any other target is obtained by integrating the flux of the density $p$ over the $j$th target,
\begin{align*}
    &\P_{x_0}(X(\tau)\in\partial\Omega_j)
    =\int_0^\infty\int_{\partial\Omega_j}D^\alpha\grad[D^{1-\alpha}p]\cdot \mathbf{n}\,\dd x\,\dd t\\
    &\quad=\frac{D(x_0)^{1-\alpha}}{\int_\Omega D^{\alpha-1}\,\dd x}\sum_{n\ge0}\frac{1}{\lambda_n}u_n^{}(x_0)\int_{\partial\Omega_j}D^\alpha\partial_{\mathbf{n}}{u}_n^{*}\,\dd x,
\end{align*}
where the second equality uses the spectral expansion in \eqref{eq:pexpand} and the dual eigenfunction relationship in \eqref{eq:dual}.
In the slow escape regime, \eqref{eq:slow} and \eqref{eq:lambdaasymptotic} imply that as $\lambda_0\to0$,
\begin{align}\label{eq:split}
    \P_{x_0}(X(\tau)\in\partial\Omega_j)
    \to\frac{\int_{\partial\Omega_j}D^\alpha\partial_{\mathbf{n}} {u}_0^{*}\,\dd x}{\sum_{j=1}^N\int_{\partial\Omega_j}D^\alpha\partial_{\mathbf{n}} {u}_0^{*}\,\dd x}.
\end{align}

To summarize, we can obtain the FPT moments and distribution, mean residence time, and splitting probabilities from knowledge of the flux into the targets of the principal backward eigenfunction ${u}_0^{*}$. We note that the limit statements in \eqref{eq:moments}-\eqref{eq:split} hold for any initial location $X(0)$ in the open set $\Omega$. However, using \eqref{eq:moments}-\eqref{eq:convdist} to approximate the moments and distribution of $\tau$ for a given $\lambda_0>0$ requires that the initial position $X(0)$ is sufficiently far from any perfectly absorbing target (we make this more precise below).

\section{\label{sec:math}Escape statistics}

We now determine the escape statistics for either perfect targets which are small (the so-called narrow escape or narrow capture problem) or imperfect targets which are small and/or weakly reactive. To facilitate the asymptotic analysis, assume that the diffusivity $D(x)$ is continuous in a neighborhood of each target and $D(x)\in[D_-,D_+]$ for all $x\in\overline{\Omega}$ with $0<D_-\le D_+<\infty$. We start with small perfect targets in a three-dimensional (3d) domain, then consider small perfect targets in a two-dimensional (2d) domain, and finally consider small and/or weakly reactive targets in a $d$-dimensional domain for $d\ge1$.

\subsection{\label{sec:3d}Perfect targets in 3d}

We begin with perfectly absorbing targets in 3d. Our setup and asymptotic analysis follows \cite{cheviakov2010asymptotic, cheviakov2011optimizing}.

Let $\Omega_0\subset\R^3$ be a bounded 3d domain with smooth boundary $\partial\Omega_0$. Without loss of generality, we choose our units of length so that $\Omega_0$ has unit volume (i.e.\ we rescale each spatial coordinate by $|\Omega_0|^{1/3}$). Let $x_1,\dots,x_{N^\text{i}}\in\Omega_0$ be $N^\text{i}\ge0$ points in the open set $\Omega_0$, which will serve as the centers of our ``interior'' targets. Let $x_{N^{\text{i}}+1},\dots,x_{N^{\text{i}}+N^\text{b}}\in\partial\Omega_0$ be $N^\text{b}\ge0$ points in the boundary of $\Omega_0$, which will serve as the centers of our ``boundary'' targets. The total number of targets is thus $N:=N^{\text{i}}+N^{\text{b}}\ge1$. 
Assume that the targets are well-separated from each other in the sense that
\begin{align*}
    |x_i-x_j|=\mathcal{O}(1)\quad\text{for all }i,j\in\{1,\dots,N\}\text{ with }i\neq j,
\end{align*}
and assume that the interior targets are well-separated from the boundary in the sense that
\begin{align*}
    \inf_{y\in\partial\Omega_0}|x_j-y|
    =\mathcal{O}(1)\quad\text{for all $j\in\{1,\dots,N^{\text{i}}\}$}.
\end{align*}

To describe the shapes of the targets, for each $n\in\{1,\dots,N\}$, let $\Omega_n'\subset\R^3$ be a simply connected open set containing the origin with an order one diameter, $\textup{diam}(\Omega_n')=\mathcal{O}(1)$. For each $n\in\{1,\dots,N^{\text{i}}\}$, define
\begin{align}\label{eq:interior}
    \Omega_{n}:=\{x\in\Omega_0:\eps^{-1}(x-x_n)\in\Omega_{n}'\},
\end{align}
where $0<\eps\ll1$. 
In words, the interior target $\Omega_n$ in \eqref{eq:interior} is a ``shrunken'' version of $\Omega_{n}'$ that is centered at $x_n$ and has $\mathcal{O}(\eps)$ diameter. 
For each $n\in\{N^{\text{i}}+1,\dots,N^{\text{i}}+N^{\text{b}}\}$, define
\begin{align}\label{eq:boundary}
    \partial\Omega_n
    :=\{x\in\partial\Omega_0:\eps^{-1}(x-x_n)\in\Omega_{n}\}.
\end{align}
In words, the boundary target $\Omega_n$ in \eqref{eq:boundary} is the intersection of $\partial\Omega_0$ with a ``shrunken'' version of $\Omega_{n}'$ that is centered at $x_n\in\partial\Omega_0$ and has $\mathcal{O}(\eps)$ diameter. 
Finally, we delete the interior targets to define the domain for our diffusion process,
\begin{align*}
    \Omega:=\Omega_0\backslash\cup_{n=1}^{N^{\text{i}}}\Omega_n.
\end{align*}

Having set up the domain and targets, we return to the spectral problem in section~\ref{sec:spectral}. Consider the principal eigenvalue $\lambda_0$ and the principal backward eigenfunction ${u}_0^{*}(x)$ in \eqref{eq:beval}. 
In the $\eps\to0$ limit, the entire boundary $\partial\Omega$ is reflecting, and thus $\lambda_0\to0$ and ${u}_0^{*}(x)\to1$ as in \eqref{eq:slow}. 
We expect that ${u}_0^{*}(x)$ rapidly changes for $x$ near a target. We thus introduce the following outer expansion which is valid away from an $\mathcal{O}(\eps)$ neighborhood of each target,
\begin{align}\label{eq:outer}
    {u}_0^{*}(x)
    \sim 1+\eps u(x)+\cdots.
\end{align}

To determine the behavior of the leading order correction $u(x)$ in \eqref{eq:outer} for $x$ near a target, we introduce inner variables around $x_j$ for fixed $j\in\{1,\dots,N\}$,
\begin{align}\label{eq:innervariables}
    y
    =\eps^{-1}(x-x_j),\quad w(y)
    ={u}_0^{*}(x_j+\eps y).
\end{align}
If we expand the inner solution as
\begin{align}\label{eq:innerexpansion}
    w\sim w_0+\cdots,
\end{align}
then $w_0$ is harmonic and vanishes on the magnified target,
\begin{align}\label{eq:3dinnerproblem}
\begin{split}
    \Delta_y w_0
    &=0,\quad y\notin\Omega',\\
    w_0
    &=0,\quad y\in\Omega',
\end{split}
\end{align}
where $\Omega'=\Omega_j'$ if $j\in\{1,\dots,N^{\text{i}}\}$ and $\Omega'=\partial\Omega_j'$ if $j\in\{N^{\text{i}}+1,\dots,N^{\text{i}}+N^{\text{b}}\}$. We note that if $j\in\{N^{\text{i}}+1,\dots,N^{\text{i}}+N^{\text{b}}\}$ (i.e.\ the $j$th target is a boundary target), then $\Omega'=\partial\Omega_j'$ is a ``flat'' 2d shape embedded in 3d, and $w_0(y)$ is only defined in half of 3d space with a reflecting boundary condition away from the magnified target in the plane containing the magnified target. However, we need write only \eqref{eq:3dinnerproblem} since symmetry enforces this reflecting boundary condition. 

If we decompose the solution to \eqref{eq:3dinnerproblem} as
\begin{align}\label{eq:3dinnerdecompose}
    w_0(y)
    =A_j(1-w_c(y))
\end{align}
then $w_c(y)$ has the following monopole decay at far-field \cite{jackson1975},
\begin{align*}
    w_c(y)
    \sim C_j/|y|\quad\text{as }|y|\to\infty,
\end{align*}
where $C_j$ is the capacitance of the shape $\Omega'$. We note that $C_j=1$ if $\Omega'$ is a unit sphere (corresponding to spherical interior targets), and $C_j=2/\pi$ if $\Omega'$ is a unit disk (corresponding to circular boundary targets). 
The matching condition is that the near-field behavior of the outer solution as $x\to x_j$ must match the far-field behavior of the inner solution as $|y|\to\infty$,
\begin{align*}
    1+\eps u(x)+\cdots
    \sim A_j\Big(1-\frac{\eps C_j}{|x-x_j|}\Big)+\cdots.
\end{align*}
which implies that $A_j=1$ and that $u$ has the following singular behavior near the $j$th target,
\begin{align}\label{eq:sing3d}
    u(x)
    \sim\frac{-C_j}{|x-x_j|}\quad\text{as }x\to x_j.
\end{align}

If the $j$th target is an interior target, then integrating over a small sphere containing the target and using the divergence theorem and the singular behavior in \eqref{eq:sing3d} yields the following flux,
\begin{align*}
    \int_{\partial\Omega_j}D^\alpha\partial_{\mathbf{n}} u\,\dd x
    \sim 
        4\pi C_j D(x_j)^\alpha\quad\text{if $j\in\{1,\dots,N^\text{i}\}$}.
\end{align*}
If the $j$th target is a boundary target, then we integrate over a hemisphere rather than a sphere and apply the same argument to obtain
\begin{align*}
    \int_{\partial\Omega_j}D^\alpha\partial_{\mathbf{n}} u\,\dd x
    \sim 
        2\pi C_j D(x_j)^\alpha\quad\text{if $j\in[N^\text{i}+1,N^\text{i}+N^\text{b}]$}.
\end{align*}
Recalling the outer expansion in \eqref{eq:outer} yields 
\begin{align}\label{eq:3dflux}
    \int_{\partial\Omega_j}D^\alpha\partial_{\mathbf{n}} {u}_0^{*}\,\dd x
    \sim 
        B_j D(x_j)^\alpha\eps\quad\text{as $\eps\to0$}, 
\end{align}
where $B_j$ is the dimensionless geometric factor,
\begin{align*}
    B_j
    :=\begin{cases}
        4\pi C_j & \text{if $j\in\{1,\dots,N^{\text{i}}\}$},\\
        2\pi C_j & \text{if $j\in\{N^\text{i}+1,\dots,N^\text{i}+N^\text{b}\}$}.
    \end{cases}
\end{align*}

By combining \eqref{eq:3dflux} with the analysis in section~\ref{sec:spectral}, we obtain the FPT moments and distribution, mean residence time, and splitting probabilities (see \eqref{eq:lambdaasymptotic}, \eqref{eq:moments}, \eqref{eq:convdist}, and \eqref{eq:split}). For instance, the mean FPT diverges according to
\begin{align}\label{eq:mfptperfect3d}
    \E_x[\tau]
    \sim\frac{\int_\Omega D^{\alpha-1}\,\dd x}{\sum_{n=1}^N (D(x_n))^\alpha B_n}\frac{1}{\eps}\quad\text{as }\eps\to0,
\end{align}
and the splitting probability limits to
\begin{align}\label{eq:splitperfect3d}
    \P_x(X(\tau)\in\partial\Omega_j)
    \to\frac{(D(x_j))^\alpha B_j}{\sum_{n=1}^N (D(x_n))^\alpha B_n}\quad\text{as }\eps\to0.
\end{align}
As noted in section~\ref{sec:spectral}, the validity of using \eqref{eq:mfptperfect3d}-\eqref{eq:splitperfect3d} requires $\eps\ll1$ and the initial condition $X(0)=x$ to be outside of an $\mathcal{O}(\eps)$ of each target, i.e.
\begin{align*}
    |x-x_n|\gg\eps\quad\text{for all }n\in\{1,\dots,N\}.
\end{align*}
We note that \eqref{eq:mfptperfect3d}-\eqref{eq:splitperfect3d} also hold if the initial position is uniformly distributed in $\Omega$ since then $X(0)$ is in an $\mathcal{O}(\eps)$ neighborhood of a target with a probability that vanishes as $\eps\to0$. That is,
\begin{align*}
    \frac{1}{|\Omega|}\int_\Omega \E_x[\tau]\,\dd x
    \sim\frac{\int_\Omega D^{\alpha-1}\,\dd x}{\sum_{n=1}^N (D(x_n))^\alpha B_n}\frac{1}{\eps}\quad\text{as }\eps\to0.
\end{align*}

\subsection{\label{sec:2d}Perfect targets in 2d}

We now consider perfectly absorbing targets in 2d. We construct the bounded domain $\Omega\subset\R^2$ and interior and boundary targets analogously to section~\ref{sec:3d} above. The analysis is also similar to section~\ref{sec:3d} above; the aspects of the analysis which are specific to 2d follow Refs.~\cite{ward10, kolokolnikov2005}.

To understand the behavior of the principal backward eigenfunction ${u}_0^{*}$ as $x\to x_j$, we define the inner variables $y$ and $w(y)$ as in \eqref{eq:innervariables} and expand the inner solution,
\begin{align*}
    w\sim \nu_j(\eps) w_0+\cdots,
\end{align*}
where $\nu_j(\eps)$ is a gauge function to be determined. Plugging the inner expansion into the backward eigenvalue problem \eqref{eq:beval} yields that $w_0$ is harmonic with an absorbing boundary condition on the target,
\begin{align}\label{eq:2dinnerproblem}
\begin{split}
    \Delta_y w_0
    &=0,\quad y\notin\Omega',\\
    w_0
    &=0,\quad y\in\Omega',
\end{split}    
\end{align}
where $\Omega'$ is defined as in \eqref{eq:3dinnerproblem} so that \eqref{eq:2dinnerproblem} holds for both interior and boundary targets.

The solution to \eqref{eq:2dinnerproblem} has the far-field behavior
\begin{align*}
    w_0(y)\sim A_j\ln(|y|/d_j)\quad\text{as }|y|\to\infty,
\end{align*}
where $d_j>0$ is the so-called logarithmic capacitance of the shape $\Omega'$. We note that $d_j=1$ if $\Omega'$ is a unit disk (corresponding to disk-shaped interior targets), and $d_j=1/4$ if $\Omega'$ is a line segment of unit length. The matching condition is then
\begin{align}\label{eq:sing2d0}
    {u}_0^{*}\sim \nu_jA_j(\ln|x-x_j|-\ln(\eps d_j))\quad\text{as }x\to x_j,
\end{align}
which yields $A_j=1$ and 
\begin{align*}
    \nu_j(\eps)=-1/\ln(\eps d_j).
\end{align*}
Thus, ${u}_0^{*}$ has the following singular behavior near the $j$th target,
\begin{align}\label{eq:sing2d}
    {u}_0^{*}
    \sim \nu_j \ln|x-x_j|\quad\text{as }x\to x_j.
\end{align}

By the 2d analog of the 3d argument used to derive \eqref{eq:3dflux} from \eqref{eq:sing3d}, the singular behavior in \eqref{eq:sing2d} yields that the flux into the $j$th target is 
\begin{align*}
    \int_{\partial\Omega_{j}}D^\alpha \partial_{\mathbf{n}} {u}_0^{*}\,\dd x
    \sim\mu_j D(x_j)^\alpha\quad\text{as }\eps\to0,
\end{align*}
where
\begin{align}\label{eq:2dflux}
    \mu_j
    =\begin{cases}
        2\pi\nu_j
        =-2\pi/\ln(\eps d_j) & \text{if }j\le N^{\text{i}},\\
        \pi\nu_j
        =-\pi/\ln(\eps d_j)& \text{if }j\ge N^{\text{i}}+1.
    \end{cases}
\end{align}

As in section~\ref{sec:3d}, combining \eqref{eq:2dflux} with the spectral analysis in section~\ref{sec:spectral} yields the FPT moments and distribution, mean residence time, and splitting probabilities (see \eqref{eq:lambdaasymptotic}, \eqref{eq:moments}, \eqref{eq:convdist}, and \eqref{eq:split}). For instance, the mean FPT diverges according to
\begin{align}\label{eq:mfptperfect2d}
    \E_x[\tau]
    \sim\frac{\int_\Omega D^{\alpha-1}\,\dd x}{\sum_{n=1}^N (D(x_n))^\alpha \mu_n}\quad\text{as }\eps\to0,
\end{align}
and the splitting probability limits to
\begin{align}\label{eq:splitperfect2d}
    \P_x(X(\tau)\in\partial\Omega_j)
    \to\frac{(D(x_j))^\alpha \mu_j}{\sum_{n=1}^N (D(x_n))^\alpha \mu_n}\quad\text{as }\eps\to0.
\end{align}

\subsection{\label{sec:kappa}Imperfect targets in any dimension}

We now consider imperfect targets in a bounded $d$-dimensional domain $\Omega\subset\R^d$. The boundary condition in \eqref{eq:Spartial} and the slow escape asymptotics in \eqref{eq:slow} yield
\begin{align}\label{eq:kappaflux}
    \begin{split}
    \int_{\partial\Omega_j}D^\alpha\partial_{\mathbf{n}}{u}_0^{*}\,\dd x
    &=\kappa_j\int_{\partial\Omega_j}D^{\alpha-1}{u}_0^{*}\,\dd x\\
    &\sim\kappa_j\int_{\partial\Omega_j}D^{\alpha-1}\,\dd x\quad\text{as }\lambda_0\to0.
    \end{split}
\end{align}
Combining \eqref{eq:kappaflux} with the spectral analysis in section~\ref{sec:spectral} yields the FPT moments and distribution, mean residence time, and splitting probabilities (see \eqref{eq:lambdaasymptotic}, \eqref{eq:moments}, \eqref{eq:convdist}, and \eqref{eq:split}). For instance, the mean FPT diverges according to
\begin{align}\label{eq:mfptkappakappa}
    \E_x[\tau]
    \sim\frac{\int_\Omega D^{\alpha-1}\,\dd y}{\sum_{n=1}^N \kappa_n\int_{\partial\Omega_n}D^{\alpha-1}\,\dd y}\quad\text{as }\lambda_0\to0,
\end{align}
and the splitting probability limits to
\begin{align}\label{eq:splitkappakappa}
    \P_x(X(\tau)\in\partial\Omega_j)
    \to\frac{\kappa_j\int_{\partial\Omega_j}D^{\alpha-1}\,\dd y}{\sum_{n=1}^N \kappa_n\int_{\partial\Omega_n}D^{\alpha-1}\,\dd y}\quad\text{as }\lambda_0\to0.
\end{align}

The principal eigenvalue $\lambda_0$ vanishes if the targets are all weakly reactive, which means that $\max_n(\kappa_n L/D_-)\to0$, where $L=|\Omega|^{1/d}$ is the length scale of the domain and $D_->0$ is a lower bound on the diffusivity $D(x)$.

The principal eigenvalue may also vanish if the targets are all small and the space dimension is $d\ge2$. In this case, we construct the domain and targets analogously to section~\ref{sec:3d} above, and \eqref{eq:kappaflux} becomes
\begin{align*}
    \int_{\partial\Omega_j}D^\alpha\partial_{\mathbf{n}}{u}_0^{*}\,\dd x
    \sim
    \eps^{d-1}\kappa_j|\partial\Omega_j'|D(x_j)^{\alpha-1}\quad\text{as }\eps\to0,
\end{align*}
where $\Omega_j'$ is the magnified target defined in section~\ref{sec:3d}. In this case, the mean FPT in \eqref{eq:mfptkappakappa} becomes
\begin{align}
    \E_x[\tau]
    &\sim\frac{\int_\Omega D^{\alpha-1}\,\dd y}{\sum_{n=1}^N\kappa_n|\partial\Omega_n'|D(x_n)^{\alpha-1}}\frac{1}{\eps^{d-1}}\quad\text{as }\eps\to0,\label{eq:mfptkappanarrow}
\end{align}
and splitting probabilities \eqref{eq:splitkappakappa} become
\begin{align}
    \P_x(X(\tau)\in\partial\Omega_j)
    &\to\frac{\kappa_j|\partial\Omega_j'|D(x_j)^{\alpha-1}}{\sum_{n=1}^N \kappa_n|\partial\Omega_n'|D(x_n)^{\alpha-1}}\quad\text{as }\eps\to0.\label{eq:splitkappanarrow}
\end{align}

\section{Higher order asymptotics}\label{sec:higherorder}

For perfect targets in 3d and 2d, we now follow the methodology of \cite{cheviakov2011optimizing, kolokolnikov2005} to determine higher order terms on the principal eigenvalue $\lambda_0$ and principal backward eigenfunction $u_0^*$ as the dimensionless target size $\eps$ vanishes. We consider only interior targets (i.e.\ $N^{\text{i}}=N\ge1$ and $N^{\text{b}}=0$) and we further suppose that $D(x)$ is continuous. In 3d, we suppose that $D(x)$ is constant in a neighborhood of each target.

These higher order corrections to $\lambda_0$ and $u_0^*$ are given in terms of the Neumann Green's function $G(x,\xi)$ of the backward operator $\mathcal{L}^*$ in \eqref{eq:backwardL}. In particular, suppose $G(x,\xi)$ satisfies
\begin{align}
\begin{split}
    \label{eq:NeumannGreens}
    \mathcal{L}^*G
    &= \frac{1}{\int_{\Omega}D^{\alpha - 1}\,\dd y} - D(x)^{1-\alpha}\delta(x-\xi),\quad x\in \Omega,\\
    \partial_{\mathbf{n}} G 
    &= 0,\quad  x\in \partial \Omega,
    \end{split}
\end{align}
with normalization
\begin{align}\label{eq:NeumannGreensnormalization}
    \int_\Omega G(x, \xi)D(x)^{\alpha-1}\,\dd x = 0.
\end{align}
In 3d, $G$ has the following singular behavior,
\begin{align}\label{eq:Gsing3d}
    G(x, \xi) \sim\frac{1}{4\pi D(\xi)^\alpha |x-\xi|} + R(\xi, \xi)\quad \text{as }x\to\xi,
\end{align}
where $R$ is the so-called regular part of $G$. In 2d, the singular behavior is logarithmic,
\begin{align}\label{eq:Gsing2d}
G(x, \xi) = \frac{-1}{2\pi D(\xi)^\alpha}\ln|x-\xi| + R(\xi, \xi) \quad \text{as }x\to\xi.
\end{align}
We derive \eqref{eq:Gsing3d} and \eqref{eq:Gsing2d} in Appendix~\ref{sec:greenssingular}.

\subsection{\label{sec:higherorder3d}Perfect targets in 3d}

Picking up from section~\ref{sec:3d}, we add the next terms to the outer expansion in \eqref{eq:outer} and the inner expansion in \eqref{eq:innerexpansion},
\begin{align}
    u_0^*(x) &\sim 1 + \eps u(x) + \eps^2 \tilde{u}(x) + \cdots,\label{eq:outerrefined}\\
    w(y) 
    &\sim w_0(y) + \eps w_1(y) + \cdots,\label{eq:innerrefined}
\end{align}
and we expand the principal eigenvalue as
\begin{align}\label{eq:lambda0refined}
    \lambda_0 &\sim \eps\Lambda_1+\eps^2\Lambda_2+\cdots.
\end{align}

If we write the singular behavior of $u$ in \eqref{eq:sing3d} in distributional form, then $u$ satisfies
\begin{align}
    \mathcal{L}^*u &= -\Lambda_1  + 4\pi\sum_j C_jD(x_j)\delta(x-x_j),\quad x\in \Omega,\label{eq:uPDE}\\
    \partial_{\mathbf{n}} u &= 0,\quad  x\in \partial \Omega.\label{eq:unoflux}
\end{align}
Integrating \eqref{eq:uPDE} and using the divergence theorem and the no flux condition \eqref{eq:unoflux} determines $\Lambda_1$,
\begin{align*}
    \Lambda_1
    =\frac{4\pi}{\int_\Omega D^{\alpha-1}\,\dd x}\sum_j C_jD(x_j)^\alpha,
\end{align*}
which agrees with the analysis in section~\ref{sec:3d}. 

The solution $u$ to \eqref{eq:uPDE}-\eqref{eq:unoflux} is then a superposition of the Neumann Green's function $G(x, \xi)$ in \eqref{eq:NeumannGreens}-\eqref{eq:NeumannGreensnormalization},
\begin{align}\label{eq:uG}
    u(x) = -4\pi\sum_{j=1}^N C_jD(x_j)^\alpha G(x, x_j).
\end{align}
Hence, \eqref{eq:uG} and \eqref{eq:Gsing3d} imply that $u$ has the following more refined singular behavior,
\begin{align*}
u(x)
&\sim
-\frac{C_j}{|x-x_j|} + A_j\quad\text{as }x\to x_j,
\end{align*}
where
\begin{align*}
    A_j
    :=-4\pi C_jD(x_j)^\alpha R(x_j, x_j)-4\pi\sum_{k \neq j} C_kD(x_k)^\alpha G(x_j, x_k).
\end{align*}
The matching condition is that the near field behavior of the outer expansion in \eqref{eq:outerrefined} as $x\to x_j$ agrees with the far-field behavior of the inner expansion in \eqref{eq:innerrefined} as $|y|\to\infty$,
\begin{align}\label{eq:matchingrefined}
    1 + \eps u + \eps^2 \tilde{u} + \cdots
    \sim w_0+\eps w_1+\cdots.
\end{align}
Hence, $w_1$ must have the following leading order far-field behavior,
\begin{align*}
    w_1(y) \sim A_j\quad\text{as }|y|\to \infty.
\end{align*}
Furthermore, assuming that $D(x)$ is constant in a neighborhood of $x_j$, it follows that $w_1$ satisfies the same problem as $w_0$ in \eqref{eq:3dinnerproblem}. We thus decompose $w_1$ as in \eqref{eq:3dinnerdecompose} to obtain the more refined far-field behavior of $w_1$,
\begin{align*}
w_1(y) = A_j(1-w_c(y))\sim A_j\Big(1-\frac{C_j}{|y|}\Big)\quad\text{as }|y|\to\infty.
\end{align*}
The matching condition \eqref{eq:matchingrefined} implies that $\tilde{u}$ must have the near-field behavior,
\begin{align}\label{eq:utildesing}
    \tilde{u}(x) \sim -\frac{A_jC_j}{|x-x_j|}\quad\text{as }x\to x_j.
\end{align}
Writing \eqref{eq:utildesing} in distributional form implies that $\tilde{u}$ satisfies
\begin{align*}
    \mathcal{L}^*\tilde{u}
    &= -\Lambda_2  - \Lambda_1u+ 4\pi\sum_j A_jC_jD(x_j)\delta(x-x_j),\quad x\in \Omega,\\
    \partial_{\mathbf{n}}\tilde{u}
    &=0,\quad x\in\partial\Omega.
\end{align*}
Integrating this equation for $\tilde{u}$, using the divergence theorem, the no flux condition on $\tilde{u}$, the representation of $u$ in \eqref{eq:uG}, and the normalization \eqref{eq:NeumannGreensnormalization} determines $\Lambda_2$,
\begin{align}\label{eq:Lambda2}
    \Lambda_2
    =-\frac{16\pi^2}{\int_{\Omega} D^{\alpha-1}\,\dd x}\sum_{j, k}C_jD(x_j)^\alpha C_kD(x_k)^\alpha \mathcal{G}_{jk},
\end{align}
where $\mathcal{G}=\{\mathcal{G}_{jk}\}_{j,k=1}^N$ is the Neumann Green's matrix,
\begin{align}\label{eq:greensmatrix}
\mathcal{G}_{jk} = \begin{cases}
    G(x_j, x_k) &\text{ if }j\neq k,\\
    R(x_j, x_j) &\text{ otherwise}.
\end{cases}
\end{align}
Plugging \eqref{eq:Lambda2} into the expansion of $\lambda_0$ in \eqref{eq:lambda0refined} yields
\begin{align*}
\lambda_0
&\sim \eps \frac{4\pi}{\int_{\Omega} D^{\alpha-1}\,\dd x}\sum_j C_jD(x_j)^{\alpha}\\
&\quad - \eps^2\frac{16\pi^2}{\int_{\Omega} D^{\alpha-1}\,\dd x}\sum_{j, k}C_jD(x_j)^\alpha C_kD(x_k)^\alpha \mathcal{G}_{jk},
\end{align*}
and plugging \eqref{eq:uG} into \eqref{eq:outerrefined} yields
\begin{align*}
    u_0^*(x) \sim 1 -4\pi\eps\sum_{j=1}^N C_jD(x_j)^\alpha G(x, x_j).
\end{align*}

\subsection{\label{sec:higherorder2d}Perfect targets in 2d}

Picking up from section~\ref{sec:2d}, we need a more detailed statement than \eqref{eq:sing2d0} and \eqref{eq:sing2d}. We still set $\nu_i(\eps) = -1/\ln(\eps d_i)$, but we leave $A_i$ in rather than setting it equal to $1$ as well as the term $\nu_iA_i\ln(\eps d_i)$ to obtain from \eqref{eq:sing2d0} that
\begin{equation}\label{eqn:sing2d_higher}
    {u}_0^{*}\sim A_i\nu_i \ln|x-x_i| + A_i\quad\text{as }x\to x_i.
\end{equation}
We expand the principal eigenvalue $\lambda_0$ and eigenfunction $u_0^*$ as
\begin{align*}
    \lambda_0(\eps)
    &\sim{{\Lambda}}(\nu)+\text{h.o.t.},\quad
    u_0^*(x,\eps)
    \sim u(x,\nu)+\text{h.o.t.},
\end{align*}
where $\nu=(\nu_1,\dots,\nu_N)$ and h.o.t.\ denotes terms which vanish faster than any power of $\nu_j$. 
Writing \eqref{eqn:sing2d_higher} in distributional form, the outer solution $u$ satisfies
\begin{align}
    \mathcal{L}^*u &= -{{\Lambda}} u -2\pi \sum_{i=1}^N A_i D(x_i) \nu_i \delta(x-x_i),\quad x\in \Omega,\label{eq:2douterpde}\\
    \partial_{\mathbf{n}} u &= 0,\quad  x\in \partial \Omega,\label{eq:2dnoflux}
\end{align}
with normalization
\begin{align*}
    (u,u)_{\alpha-1}
    =\int_\Omega u^2D^{\alpha-1}\,\dd x = \int_\Omega D^{\alpha-1}\,\dd x.
\end{align*}
To solve \eqref{eq:2douterpde}-\eqref{eq:2dnoflux}, introduce the Helmholtz Green's function $G_h(x,\xi,{{\Lambda}})$ as the solution to
\begin{equation}
    \begin{aligned}
        \mathcal{L}^*G_h &= -{{\Lambda}}G_h-\frac{1}{D(\xi)^{\alpha-1}}\delta(x-\xi),\quad x\in \Omega,\\
        \partial_{\mathbf{n}} G_h 
        &= 0,\quad  x\in \partial \Omega.
    \end{aligned}
    \label{eqn:2dhigher_greens}
\end{equation}
The outer solution $u$ is then a superposition of Helmholtz Green's functions,
\begin{equation}
    u(x)=-2\pi \sum_{j=1}^N A_j D(x_j)^\alpha \nu_j G_h(x, x_j, {{\Lambda}}).
    \label{eqn:2dhigher_uGreens}
\end{equation}

Expand the Helmholtz Green's function as ${{\Lambda}}$ vanishes,
\begin{align}\label{eq:helmexpansion}
    G_h(x, \xi, {{\Lambda}})
    = \frac{1}{{{\Lambda}}}G_0(x, \xi) + G(x, \xi) + \cdots. 
\end{align}
Plugging this expansion into \eqref{eqn:2dhigher_greens} yields the leading order problem,
\begin{align*}
    \mathcal{L}^*G_0
    = 0, \quad \partial_{\mathbf{n}}G_0 = 0,
\end{align*}
which implies that $G_0$ is constant in space. The next order problem is
\begin{align}\label{eq:GG0}
    \mathcal{L}^*G
    = -G_0-\frac{1}{D(\xi)^{\alpha-1}}\delta(x-\xi), \quad \partial_{\mathbf{n}}G = 0.
\end{align}
Multiplying by $D^{\alpha-1}$, integrating over $\Omega$, and using the divergence theorem and the no flux boundary condition determines the constant $G_0$,
\begin{align*}
    G_0 = \frac{-1}{\int_\Omega D^{\alpha-1}\,\dd x}.
\end{align*}
Hence, the function $G(x,\xi)$ satisfying \eqref{eq:GG0} is the Neumann Green's function satisfying \eqref{eq:NeumannGreens}-\eqref{eq:NeumannGreensnormalization} with the singular behavior in \eqref{eq:Gsing2d}. Therefore, combining \eqref{eqn:2dhigher_uGreens} with the expansion \eqref{eq:helmexpansion} yields the following near-field behavior of $u(x)$ as $x\to x_i$,
\begin{align}\begin{split}
    \label{eq:sing2d_higherrefined}
    u(x)
    &\sim A_i\nu_i\ln|x-x_i|+\frac{2\pi}{{{\Lambda}}\int_\Omega D^{\alpha-1}\,\dd y}\sum_j A_j D(x_j)^\alpha\nu_j\\
    &\quad-2\pi\sum_j A_j D(x_j)^\alpha\nu_j\mathcal{G}_{ij}\quad\text{as }x\to x_i,
    \end{split}
\end{align}
where $\mathcal{G}$ is the Neumann Green's matrix in \eqref{eq:greensmatrix}.

We now equate \eqref{eqn:sing2d_higher} and \eqref{eq:sing2d_higherrefined} to obtain a system of linear algebraic equations which determine $A_1,\dots,A_N$. Noting that the logarithmic terms in \eqref{eqn:sing2d_higher} and \eqref{eq:sing2d_higherrefined} agree, we obtain
\begin{align*}
    &A_i+2\pi \sum_{j} A_jD(x_j)^{\alpha}\nu_j\mathcal{G}_{ij}\\
    &\quad= \frac{2\pi}{{{\Lambda}}\int_\Omega D^{\alpha-1}\,\dd x} \sum_{j} A_jD(x_j)^{\alpha}\nu_j,\quad i=1,\dots,N.
\end{align*} 
Setting $\textbf{1}$ to be the vertical vector of all 1s, $\mathcal{A}$ to be the vertical vector with entries $A_i$, and $\mathcal{V}$ to be a diagonal matrix with entries $D(x_i)^\alpha \nu_i$, the system becomes
$$
(I+2\pi\mathcal{G}\mathcal{V})\mathcal{A} = \frac{2\pi}{{{\Lambda}}\int_\Omega D^{\alpha-1}\,\dd x} \textbf{1}\textbf{1}^\top\mathcal{V}\mathcal{A}.
$$
Multiplying by ${{\Lambda}}$ and using that $(I+2\pi\mathcal{G}\mathcal{V})^{-1} \sim I-2\pi\mathcal{G}\mathcal{V}$ as $\eps\to0$, we obtain
$$
{{\Lambda}}\mathcal{A} = \frac{2\pi}{\int_\Omega D^{\alpha-1}\,\dd x} (I-2\pi\mathcal{G}\mathcal{V})\textbf{1}\textbf{1}^\top\mathcal{V}\mathcal{A}.
$$
From here, observe that ${{\Lambda}}, \mathcal{A}$ is an eigenvalue and eigenvector pair of the matrix
\begin{align*}
    \frac{2\pi}{\int_\Omega D^{\alpha-1}\,\dd x} (I-2\pi\mathcal{G}\mathcal{V})\textbf{1}\textbf{1}^\top\mathcal{V},
\end{align*}
which is a rank-one matrix with eigenvalue
\begin{align*}
    {{\Lambda}}
    &= \frac{2\pi}{\int_\Omega D^{\alpha-1}\,\dd x} \textbf{1}^\top\mathcal{V}(I-2\pi\mathcal{G}\mathcal{V})\textbf{1}\\
    &= \frac{2\pi}{\int_\Omega D^{\alpha-1}\,\dd x}\bigg[\sum_i D(x_i)^\alpha \nu_i\\
    &\qquad\qquad\qquad\qquad-2\pi \sum_{i,  j} \mathcal{G}_{ij}D(x_j)^\alpha\nu_jD(x_i)^\alpha \nu_i\bigg],
\end{align*}
and eigenvector $\mathcal{A} = (I-2\pi\mathcal{G}\mathcal{V})\textbf{1}$, and therefore 
\begin{align*}
    A_j&\sim 1-2\pi \sum_{j=1}^N \mathcal{G}_{ij}D(x_j)^\alpha\nu_j.
\end{align*}

\section{Stochastic simulations}\label{sec:sims}

We now compare our analytical theory to stochastic simulations. We first present results for perfect targets in 3d, then perfect targets in 2d, and finally imperfect targets in 2d. 

\subsection{Perfect targets in 3d}

Consider the 3d unit cube domain,
\begin{align*}
    \Omega
    =\{x=(y_1,y_2,y_3)\in(0,1)^3\}\subset\R^3,
\end{align*}
with $N=2$ perfectly absorbing, disk-shaped targets with common radius $a\ll1$ embedded at the center of the left and right ``walls,''
\begin{align*}
     \partial\Omega_1
     &=\{x=(0,y_2,y_3)\in\R^3:(y_2-\tfrac{1}{2})^2+(y_3-\tfrac{1}{2})^2<a^2\},\\
     \partial\Omega_2
     &=\{x=(1,y_2,y_3)\in\R^3:(y_2-\tfrac{1}{2})^2+(y_3-\tfrac{1}{2})^2<a^2\}.
\end{align*}
Suppose the diffusivity is the following linear function of the single coordinate $y_1$ of the 3d vector $x=(y_1,y_2,y_3)$,
\begin{align}\label{eq:simD}
    D(x)
    =b_0+(b_1-b_0)y_1,
\end{align}
where $b_0=0.1\ll b_1=10$. Suppose the searcher starts in the center of the domain, $X(0)=(1/2,1/2,1/2)\in\Omega$.

\begin{figure}
\centering
\includegraphics[width=1\linewidth]{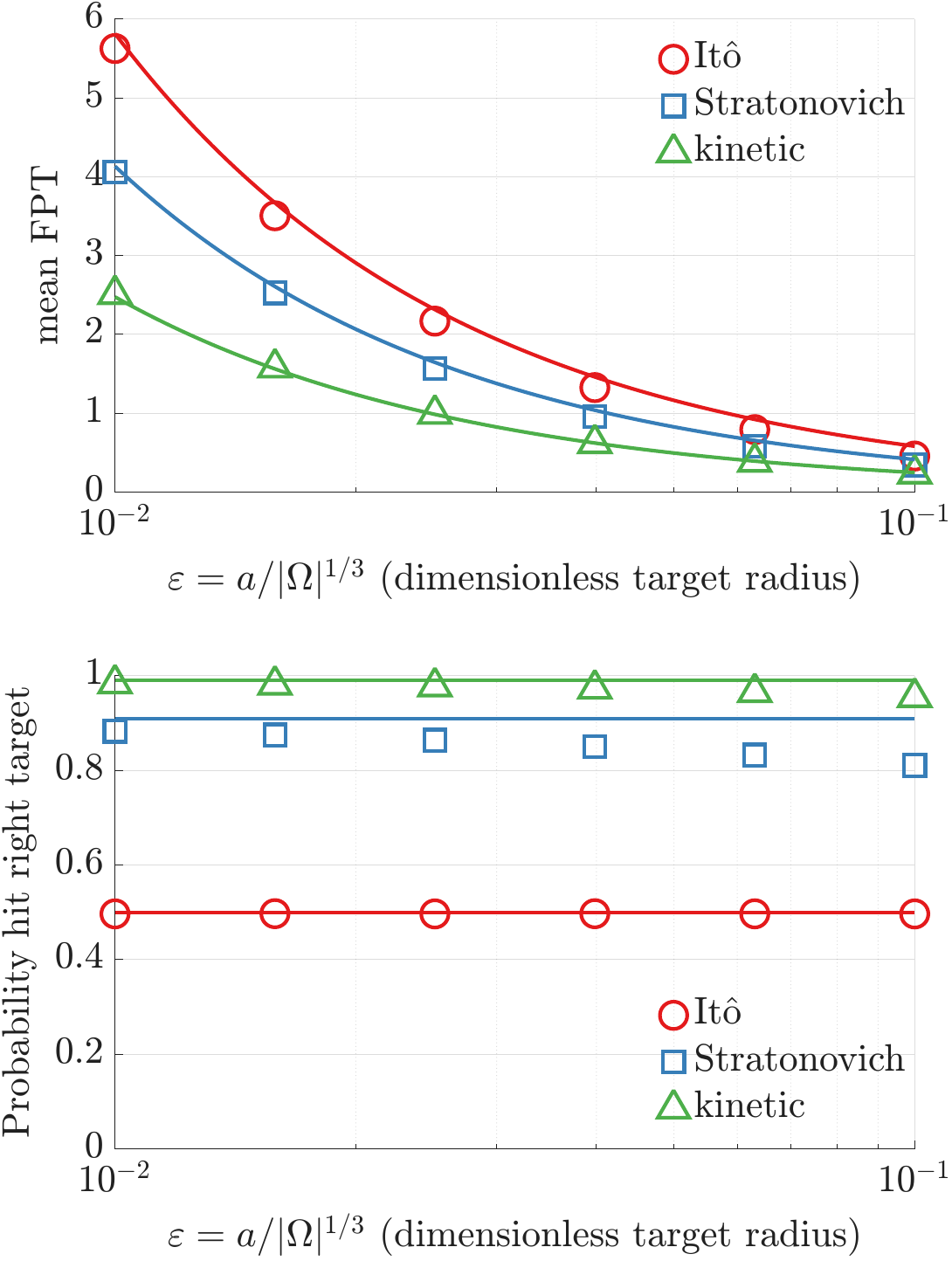}
\caption{Perfect targets in 3d for the It\^{o} ($\alpha=0$), Stratonovich ($\alpha=1/2$), and kinetic interpretations ($\alpha=1$). The markers are computed from stochastic simulations, and the curves are the formulas in \eqref{eq:mfptperfect3d} and \eqref{eq:splitperfect3d}.}
\label{fig:sim3da}
\end{figure}

Figure~\ref{fig:sim3da} shows close agreement between stochastic simulations (markers) and the formulas (curves) for the mean FPT in \eqref{eq:mfptperfect3d} and splitting probability in \eqref{eq:splitperfect3d} for the It\^{o} ($\alpha=0$), Stratonovich ($\alpha=1/2$), and kinetic interpretations ($\alpha=1$). Each marker in this figure is computed from $5\times10^6$ independent simulations of the heterogeneous diffusion process. The diffusive paths are simulated using the Euler-Maruyama method after converting the driftless stochastic differential equation in \eqref{eq:sde0} into a stochastic differential equation of It\^{o} form (see \eqref{eq:convert}).

\subsection{Perfect targets in 2d}

Consider the 2d unit square domain,
\begin{align*}
    \Omega
    =\{x=(y_1,y_2)\in(0,1)^2\}\subset\R^2,
\end{align*}
with $N=2$ perfectly absorbing targets with common radius $a\ll1$ embedded at the center of the left and right ``walls,''
\begin{align*}
     \partial\Omega_1
     &=\{x=(0,y_2)\in\R^2:(y_2-\tfrac{1}{2})^2<a^2\},\\
     \partial\Omega_2
     &=\{x=(1,y_2)\in\R^2:(y_2-\tfrac{1}{2})^2<a^2\}.
\end{align*}
That is, the targets are line segments of length $2a\ll1$. Suppose the diffusivity is given by \eqref{eq:simD} with $b_0=0.1\ll b_1=10$. Suppose the searcher starts in the center of the domain, $X(0)=(1/2,1/2)\in\Omega$.

\begin{figure}
\centering
\includegraphics[width=1\linewidth]{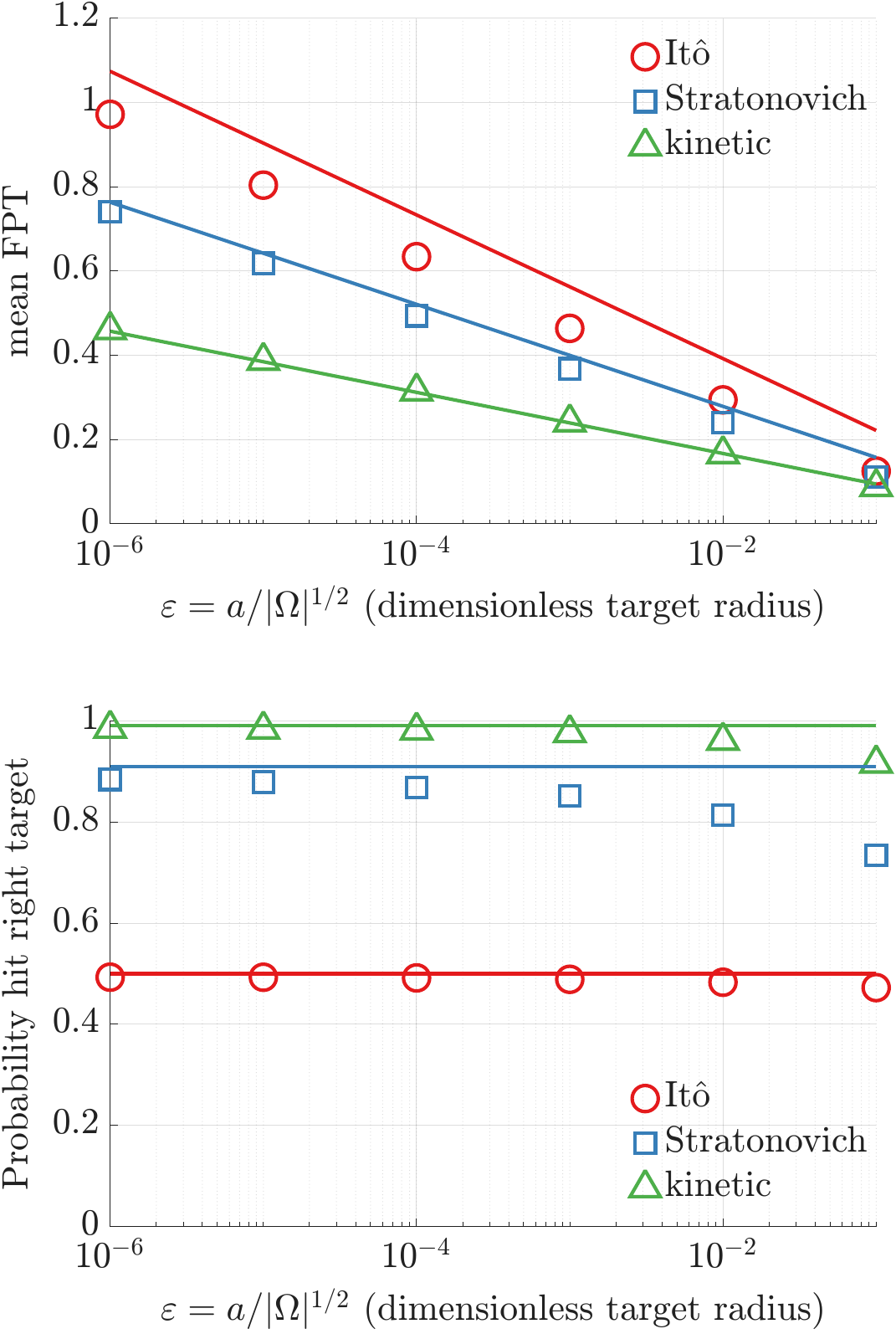}
\caption{Perfect targets in 2d for the It\^{o} ($\alpha=0$), Stratonovich ($\alpha=1/2$), and kinetic interpretations ($\alpha=1$). The markers are computed from stochastic simulations, and the curves are the formulas in \eqref{eq:mfptperfect2d} and \eqref{eq:splitperfect2d}.}
\label{fig:sim2da}
\end{figure}

Figure~\ref{fig:sim2da} compares stochastic simulations (markers) to the formulas (curves) for the mean FPT in \eqref{eq:mfptperfect2d} and splitting probability in \eqref{eq:splitperfect2d}. Each marker in this figure is computed from $10^6$ independent simulations of the heterogeneous diffusion process. The diffusive paths are simulated as described above.

\begin{figure}
\centering
\includegraphics[width=1\linewidth]{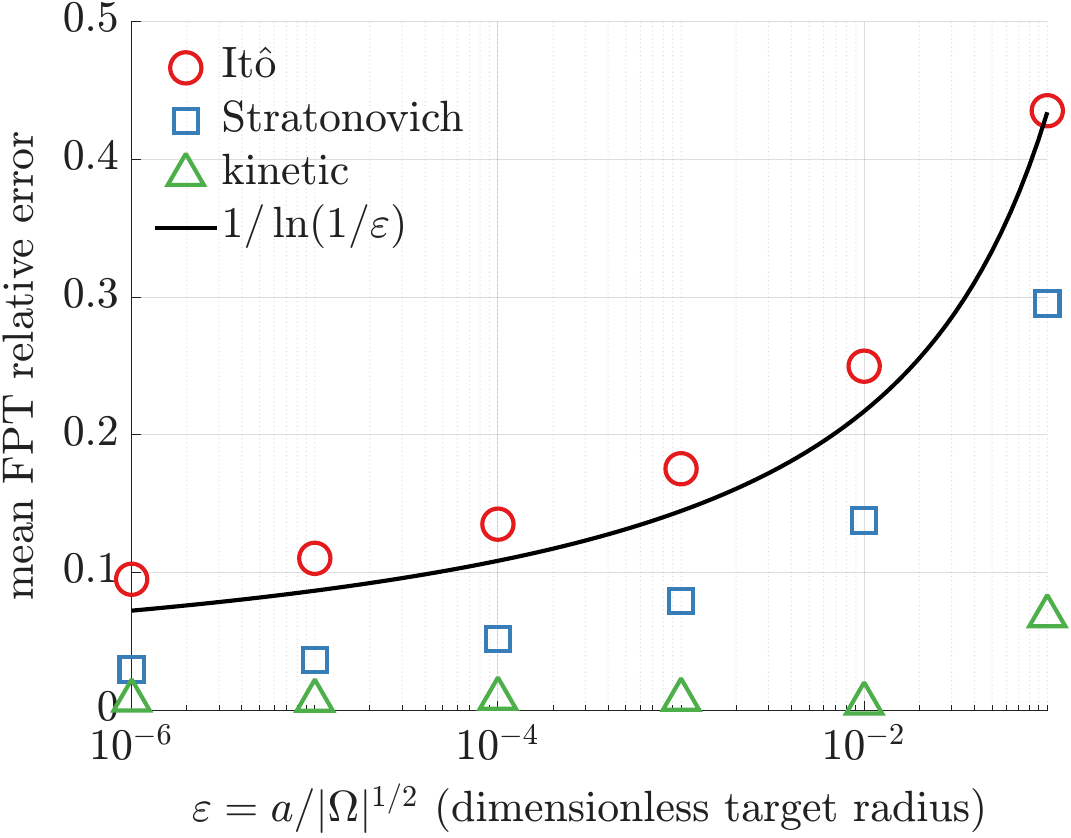}
\caption{Relative mean FPT error for perfect targets in 2d for the It\^{o} ($\alpha=0$), Stratonovich ($\alpha=1/2$), and kinetic interpretations ($\alpha=1$)}
\label{fig:sim2daError}
\end{figure}

Since the convergence between the simulations and the mean FPT theory as $\eps$ vanishes is not visually evident in this case of 2d perfect targets, Figure~\ref{fig:sim2daError} plots the relative error between the simulations and theory. This figure confirms that the mean FPT relative error decays as $1/\ln(1/\eps)$ as $\eps\to0$, which agrees with higher order estimates obtained for a constant diffusivity \cite{ward10}. Interestingly, this figure shows that the error is much smaller for the kinetic ($\alpha=1$) interpretation of the multiplicative noise compared to the It\^{o} ($\alpha=0$) interpretation.

\subsection{Imperfect targets in 2d}

\begin{figure}
\centering
\includegraphics[width=1\linewidth]{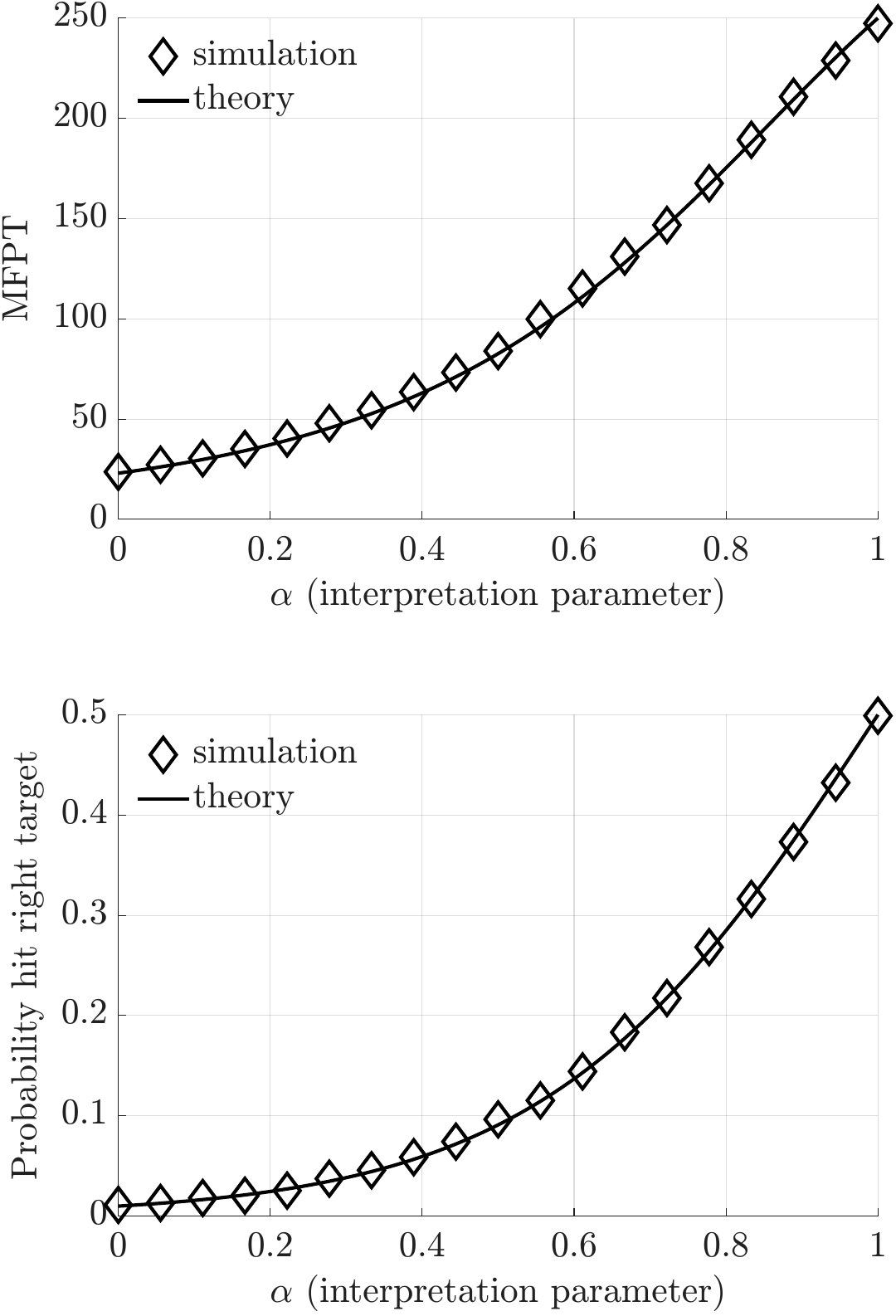}
\caption{Imperfect targets in 2d. The markers are computed from stochastic simulations, and the curves are the formulas in \eqref{eq:mfptkappanarrow} and \eqref{eq:splitkappanarrow}.}
\label{fig:sim2d}
\end{figure}

Under the same 2d setup as the preceding section, suppose now that the targets are imperfect with a common unit reactivity, $\kappa_1=\kappa_2=1$. Figure~\ref{fig:sim2d} shows close agreement between stochastic simulations (markers) and the formulas (curves) in \eqref{eq:mfptkappanarrow} and \eqref{eq:splitkappanarrow} as the interpretation parameter $\alpha$ varies from $\alpha=0$ to $\alpha=1$. Each marker in this plot is computed from $10^4$ independent simulations of the heterogeneous diffusion process with $\eps=10^{-3}$.

The diffusion process is simulated as above, except that absorption at the imperfect targets is simulated using the method of Erban and Chapman \cite{erban2007reactive}. Specifically, if the diffusion process hits target $i\in\{1,2\}$, then it is absorbed with probability $\kappa_i\sqrt{\pi\Delta t/D(x_i)}$ and is otherwise reflected.

\section{Comparing It\^{o}, Stratonovich, and kinetic}\label{sec:implications}

We now use the results above to compare stochastic search with either the It\^{o}, Stratonovich, or kinetic definitions of heterogeneous diffusion.

\subsection{\label{sec:prelim}Preliminaries}
To aid the explanations below, we first recall some facts about a heterogeneous diffusion process $X=\{X(t)\}_{t\ge0}$ with multiplicative noise parameter $\alpha\in[0,1]$. If $X$ diffuses in a bounded domain $\Omega\subset\R^d$ with reflecting boundaries, then its steady-state probability density is
\begin{align}\label{eq:steady}
    p(x)
    =\mathcal{N}D(x)^{\alpha-1},
\end{align}
where the normalization constant $\mathcal{N}=(\int_\Omega D^{\alpha-1}\,\dd x)^{-1}$ is such that $\int_\Omega p\,\dd x=1$. Furthermore, though $X$ has multiplicative noise parameter $\alpha\in[0,1]$, we can write its governing stochastic differential equation in the following It\^{o} form with a drift term that biases $X$ to move up the gradient of the diffusivity,
\begin{align}\label{eq:convert}
    \dd X
    =\alpha \grad D(X)\,\dd t+\sqrt{2D(X)}\cdot\dd W,
\end{align}
assuming that the diffusivity $D$ is a smooth function of space (we use ``$\sqrt{2D(X)}\cdot\dd W$'' in \eqref{eq:convert} to denote multiplicative noise with It\^{o} interpretation).

The conversion to It\^{o} form in \eqref{eq:convert} has three important implications.  
First, if $\alpha=0$, then the drift term in \eqref{eq:convert} vanishes, and it follows that the path $X=\{X(t)\}_{t\ge0}$ is a random time change of unbiased Brownian motion. In particular, if $\alpha=0$, then $X$ is an unbiased Brownian motion that moves slower in regions of space where the diffusivity is smaller. This fact is reflected in the density in \eqref{eq:steady}, which is inversely proportional to $D(x)$ if $\alpha=0$ (i.e.\ $X$ is more likely to be found where $D$ is small since $X$ moves slowly there). 
Second, if $\alpha>0$, then the drift term in \eqref{eq:convert} biases $X$ toward regions where the diffusivity is large. This fact is also reflected in \eqref{eq:steady}, since the density becomes more uniform for $\alpha>0$. Third, if $\alpha=1$, then the drift in \eqref{eq:convert} toward regions with large diffusivity balances the faster motion in those regions so that the density in \eqref{eq:steady} is exactly uniform, $p(x)=1/|\Omega|$.

\subsection{FPT to perfect targets}\label{eq:perfectexplore}

We start with perfect targets. The difference in the 3d versus 2d results in sections~\ref{sec:3d}-\ref{sec:2d} involves merely a $1/\eps$ versus a $\ln(1/\eps)$ divergence. Furthermore, interior targets versus boundary targets entails only an additional factor of 2, and different sizes and shapes of targets merely entails different values of the associated capacitance. Hence, to simplify the exposition, suppose there is a single disk-shaped target with small radius $a>0$ centered at a point $x_1\in\partial\Omega$ on the boundary of a 3d domain $\Omega$. Using \eqref{eq:mfptperfect3d}, the MFPT diverges according to
\begin{align}\label{eq:mfptperfectexplore}
    \E[\tau]
    \sim\frac{1}{4a}\frac{\int_{\Omega} D^{\alpha-1}\,\dd x}{ (D(x_1))^{\alpha}}\quad\text{as }a/|\Omega|^{1/3}\to0.
\end{align}

Setting $\alpha=0$, $\alpha=1/2$, or $\alpha=1$ in the general $\alpha$-dependent expression for the MFPT in \eqref{eq:mfptperfectexplore} yields the following It\^{o}, Stratonovich, and kinetic MFPTs,
\begin{align}
    T_{\text{ito}}
    &:=\frac{1}{4 a}\int_\Omega 1/D\,\dd x,\label{eq:Tito}\\
    T_{\text{strat}}
    &:=\frac{1}{4 a}\frac{\int_\Omega 1/\sqrt{D}\,\dd x}{\sqrt{D(x_1)}},\label{eq:Tstrat}\\
    T_{\text{kin}}
    &:=\frac{1}{4 a}\frac{|\Omega|}{D(x_1)}.\label{eq:Tkin}
\end{align}
A prominent feature of \eqref{eq:Tito}-\eqref{eq:Tkin} is that (i) $T_{\text{ito}}$ depends on the diffusivity only through the global average of $1/D$, whereas (ii) $T_{\text{kin}}$ depends only on the local diffusivity at each target. Naturally, $T_{\text{strat}}$ is between these two extremes. Points (i) and (ii) can be understood in terms of the properties of heterogeneous diffusion reviewed in section~\ref{sec:prelim}. In particular, a driftless It\^{o} path is merely a time-changed, unbiased Brownian motion, and thus the MFPT is scaled by the time scale of domain exploration and is agnostic to whether the targets are located in regions of high or low diffusivity. In contrast, a kinetic interpretation of heterogeneous diffusion entails a strong drift toward regions of high diffusivity. Thus, $T_{\text{kin}}$ is minimized by placing the targets in regions of high diffusivity.

Another interesting feature of \eqref{eq:Tito}-\eqref{eq:Tkin} is that merely varying the value of $\alpha\in[0,1]$ can greatly affect the FPT (without changing the diffusivity or any other parameters). In fact, it is possible to have $T_{\text{ito}}\gg T_{\text{kin}}$ or $T_{\text{ito}}\ll T_{\text{kin}}$, depending on the locations of the targets. 

Another notable feature of \eqref{eq:Tito}-\eqref{eq:Tkin} is that if $\alpha>0$, then the target locations affect the FPT at leading order in the small target limit. In contrast, if the diffusivity is constant in space, then it is well-known that the leading order FPT does not depend on target locations \cite{cheviakov2010asymptotic}.

\subsection{FPT to imperfect targets}

We now consider imperfect targets as in section~\ref{sec:kappa}. As in section~\ref{eq:perfectexplore}, we consider a single disk-shaped target with small radius $a>0$ and reactivity $\kappa>0$ centered at point $x_1\in\partial\Omega$ on the boundary of a 3d domain $\Omega$. Using \eqref{eq:mfptkappanarrow}, the MFPT diverges according to
\begin{align}\label{eq:mfptimperfectexplore}
    \E[\tau]
    &\sim\frac{(D(x_1))^{1-\alpha}}{\kappa\pi a^2} \int_{\Omega} \frac{1}{D^{1-\alpha}}\,\dd x\quad\text{as }\frac{a}{|\Omega|^{1/3}}\to0.
\end{align}

Setting $\alpha=0$, $\alpha=1/2$, or $\alpha=1$ in the general $\alpha$-dependent expression for the MFPT in \eqref{eq:mfptimperfectexplore} yields the following It\^{o}, Stratonovich, and kinetic MFPTs,
\begin{align}
    T_{\text{ito}}^\kappa
    &:=\frac{D(x_1)}{\kappa\pi a^2}\int_\Omega \frac{1}{D}\,\dd x,\label{eq:Titokappa}\\
    T_{\text{strat}}^\kappa
    &:=\frac{\sqrt{D(x_1)}}{\kappa\pi a^2}\int_\Omega \frac{1}{\sqrt{D}}\,\dd x,\label{eq:Tstratkappa}\\
    T_{\text{kin}}^\kappa
    &:=\frac{|\Omega|}{\kappa\pi a^2}.\label{eq:Tkinkappa}
\end{align}
Perhaps the most interesting feature of \eqref{eq:Titokappa}-\eqref{eq:Tkinkappa} is that a high diffusivity at the targets increases $T_{\text{ito}}^\kappa$ and $T_{\text{strat}}^\kappa$ and does not affect $T_{\text{kin}}^\kappa$, whereas \eqref{eq:Tito}-\eqref{eq:Tkin} shows that a high diffusivity at the targets decreases $T_{\text{strat}}$ and $T_{\text{kin}}$ and does not affect $T_{\text{ito}}$. This discrepancy can be understood by noting that a higher diffusivity at the targets has the two opposing effects of (a) biasing the searcher to move toward the targets if $\alpha>0$, which decreases the FPT, and (b) decreasing the time spent near a target, which increases the FPT if the target is imperfect since absorption at an imperfect target requires accumulating enough local time at the target (see \eqref{eq:taukappa}). Effect (b) is irrelevant for perfect targets, and thus effect (a) explains why a high diffusivity at the targets decreases $T_{\text{strat}}$ and $T_{\text{kin}}$. Furthermore, the diffusivity at the targets does not affect $T_{\text{ito}}$ since effects (a) and (b) are both irrelevant for perfect targets with $\alpha=0$. For imperfect targets, only effect (b) is relevant if $\alpha=0$, which explains why $T_{\text{ito}}^\kappa$ increases if the diffusivity is high at the targets. If $\alpha>0$ for imperfect targets, then effects (a) and (b) are both relevant, with effect (a) becoming stronger with larger values of $\alpha$. In fact, if $\alpha=1$, then effects (a) and (b) perfectly balance to yield the result that $T_{\text{kin}}^\kappa$ is independent of the diffusivity.

\subsection{Splitting probabilities}

To compare how the It\^{o}, Stratonovich, and kinetic interpretations affect splitting probabilities, suppose there are $N\ge2$ targets located at $x_1,\dots,x_N$ in a 3d or 2d domain. To simplify the exposition, suppose the targets all have the same size and shape and that they are all interior targets or all boundary targets. 

If the targets are perfectly absorbing, then \eqref{eq:splitperfect3d} (or \eqref{eq:splitperfect2d}) implies that the probability that the searcher finds the $j$th target satisfies
\begin{align}\label{eq:Pperfectexplore}
    \P_x(X(\tau)\in\partial\Omega_j)
    \to\frac{(D(x_j))^\alpha}{\sum_{n=1}^N(D(x_n))^\alpha}\quad\text{as }\frac{a}{|\Omega|^{1/3}}\to0.
\end{align}
Setting $\alpha=0$, $\alpha=1/2$, or $\alpha=1$ in the $\alpha$-dependent splitting probability for perfect targets in \eqref{eq:Pperfectexplore} yields the probability that the searcher finds the $j$th target for It\^{o}, Stratonovich, and kinetic interpretations,
\begin{align}
    P_{\text{ito}}
    &:=\frac{1}{N},\label{eq:Pito}\\
    P_{\text{strat}}
    &:=\frac{\sqrt{D(x_j)}}{\sum_{n=1}^N \sqrt{D(x_n)}},\label{eq:Pstrat}\\
    P_{\text{kin}}
    &:=\frac{D(x_j)}{\sum_{n=1}^N D(x_n)}.\label{eq:Pkin}
\end{align}
Notice that \eqref{eq:Pito} implies that $P_{\text{ito}}$ is independent of the diffusivity. This point accords with the fact that a driftless It\^{o} diffusion is a time-change of an unbiased Brownian motion, and mere time changes cannot affect which target the searcher ultimately finds. This point also agrees with the case of a diffusivity that is constant in space. In contrast, \eqref{eq:Pstrat}-\eqref{eq:Pkin} show that the diffusivity affects $P_{\text{strat}}$ and $P_{\text{kin}}$, but only via the value of the diffusivity at the targets. This result reflects the fact that the searcher is biased toward regions of higher diffusion if $\alpha>0$, and therefore the searcher is more likely to find targets in regions of high diffusivity. Naturally, this effect is stronger for $P_{\text{kin}}$ ($\alpha=1$) than $P_{\text{strat}}$ ($\alpha=1/2$).

If the targets are imperfect, then  \eqref{eq:splitkappanarrow} yields
\begin{align}\label{eq:Pimperfectexplore}
    \P_x(X(\tau)\in\partial\Omega_j)
    \to\frac{(D(x_j))^{\alpha-1}}{\sum_{n=1}^N(D(x_n))^{\alpha-1}}\quad\text{as }\frac{a}{|\Omega|^{1/3}}\to0.
\end{align}
Setting $\alpha=0$, $\alpha=1/2$, or $\alpha=1$ in \eqref{eq:Pimperfectexplore} yields the following probability that the searcher finds the $j$th target for It\^{o}, Stratonovich, and kinetic interpretations with imperfect targets,
\begin{align}
    P_{\text{ito}}^\kappa
    &:=\frac{1/D(x_j)}{\sum_{n=1}^N 1/D(x_n)}, \label{eq:Pitokappa}\\
    P_{\text{strat}}^\kappa
    &:=\frac{1/\sqrt{D(x_j)}}{\sum_{n=1}^N 1/\sqrt{D(x_n)}}, \label{eq:Pstratkappa}\\
    P_{\text{kin}}^\kappa
    &:=\frac{1}{N}. \label{eq:Pkinkappa}
\end{align}
Interestingly,
\begin{align*}
    P_{\text{kin}}^\kappa=P_{\text{ito}}=1/N    
\end{align*}
Notice also that for the Stratonovich interpretation, the searcher is most likely to find a target at the smallest diffusivity if the targets are imperfect, whereas the opposite is true for the Stratonovich interpretation with perfect targets.

\section{Discussion}

In this paper, we investigated how a spatially dependent diffusivity and its interpretation parameter $\alpha\in[0,1]$ affect stochastic search. We derived general asymptotic formulas for all the moments of the FPT, the limiting FPT distribution, the mean residence time in any subset of the domain, and the splitting probability. Our results are for a general space-dependent diffusivity $D(x)$ in general spatial domains with multiple perfect or imperfect targets located in the interior and/or the boundary of the domain. Our results are valid in the parameter regime in which the diffusive searcher tends to wander around the entire confining domain before finding a target. This ``slow escape'' regime holds if the targets are small (i.e.\ the so-called narrow escape or narrow capture problem) and/or weakly reactive.

As important prior work, Godec and Metzler investigated how space-dependent diffusivity affects stochastic search for $\alpha=1$ (kinetic interpretation) for a single perfect target in the center of a radially symmetric domain with a radially symmetric, piecewise constant diffusivity with two possible values \cite{godec2015optimization, godec2016first}. The symmetry makes the problem one-dimensional, so that the domain is an interval,
\begin{align*}
    \Omega
    =(a,R)\subset\R,
\end{align*}
with absorption at the inner radius $r=a$ and reflection at the outer radius $r=R>a$, and the diffusivity is either $D(r)=D_1$ if $r<r_{\text{interface}}$ or $D(r)=D_2$ if $r>r_{\text{interface}}$ for some $r_{\text{interface}}\in(a,R)$. These authors obtained the very interesting and counterintuitive result that if the searcher starts in the inner region (i.e.\ $X(0)<r_{\text{interface}}$), then the mean FPT is independent of the outer region diffusivity $D_2$. This prior exact result accords with our general asymptotic result that the mean FPT (and indeed, all the moments of the FPT and its probability distribution) only depends on the diffusivity at the (perfect) target for the kinetic interpretation.

Vaccario, Antoine, and Talbot \cite{vaccario2015} generalized the radially symmetric framework of Ref.~\cite{godec2015optimization} by allowing the interpretation parameter to vary in the interval $\alpha\in[0,1]$ and computing mean residence times. Ref.~\cite{vaccario2015} found similar counterintuitive results to Ref.~\cite{godec2015optimization} for the kinetic interpretation of multiplicative noise. Interestingly, Ref.~\cite{vaccario2015} also found that the mean residence time in the region containing the target is independent of the diffusivity in the outer region for any $\alpha\in[0,1]$. This exact result accords with our general asymptotic result that the mean residence time $\E[\tau_{\Omega'}]$ in any region $\Omega'\subset\Omega$ depends only on the diffusivity at the targets and an average of the diffusivity in $\Omega'$ (i.e.\ $\int_{\Omega'}D^{\alpha-1}\,\dd x$). Indeed, combining \eqref{eq:residence} with \eqref{eq:moments} implies that the mean residence time and the mean FPT have the following relation in the slow escape regime,
\begin{align*}
    \E[\tau_{\Omega'}]
    \sim\frac{\int_{\Omega'}D^{\alpha-1}\,\dd x}{\int_{\Omega}D^{\alpha-1}\,\dd x}\E[\tau].
\end{align*}
Hence, our mean FPT $\E[\tau]$ results (namely, \eqref{eq:mfptperfect3d}, \eqref{eq:mfptperfect2d}, and \eqref{eq:mfptkappanarrow}) generalize to mean residence time $\E[\tau_{\Omega'}]$ results by simply substituting the integral of $D^{\alpha-1}$ over $\Omega$ to an integral over $\Omega'$.

As additional related work, Grebenkov assumed an It\^{o} interpretation and studied the mean FPT to a single perfect target on the boundary of a general 2d domain with a general diffusivity $D(x)$ \cite{grebenkov2016}. Based on an innovative application of the Riemann mapping theorem, Grebenkov used a conformal map to obtain an exact MFPT formula, which agrees with our result in \eqref{eq:mfptperfect2d} for $\alpha=0$ if there is a single perfect target on the boundary of a 2d domain.


\subsubsection*{Acknowledgments}
SDL was supported by the National Science Foundation (DMS-2325258).


\appendix
\section{\label{sec:greenssingular}Green's function singular behavior}

For the Neumann Green's function $G(x,\xi)$ satisfying \eqref{eq:NeumannGreens}, we now derive the singular behavior in \eqref{eq:Gsing3d} for 3d and \eqref{eq:Gsing2d} for 2d. 

Starting with 3d, suppose a function $u(x)$ satisfies
\begin{align}\label{eq:pdeu}
    \nabla\cdot[D(x)^\alpha\nabla u]
    =f(x)+A\delta(x-x_0),\quad x\in\Omega,
\end{align}
with the singularity,
\begin{align}\label{eq:usingderivation}
    u(x)
    \sim\frac{B}{|x-x_0|}\quad\text{as }x\to x_0\in\Omega\subset\R^3.
\end{align}
Let $\Omega_\eps\subset\Omega$ be a sphere of small radius $\eps>0$ centered on $x_0$,
\begin{align*}
    \Omega_\eps=\{x\in\Omega:|x-x_0|\le\eps\}.
\end{align*}
Integrating \eqref{eq:pdeu} over $\Omega_\eps$ and using the divergence theorem and the singularity in \eqref{eq:usingderivation} gives
\begin{align*}
    A\sim\int_{\Omega_\eps}f(x)\,\dd x+A
    &=\int_{\partial\Omega_\eps}D(x)^\alpha\nabla u\cdot \mathbf{n}\,\dd S\\
    &\sim -4\pi D(x_0)^\alpha B\quad\text{as }\eps\to0,
\end{align*}
and thus
\begin{align*}
    B
    =\frac{-A}{4\pi D(x_0)^\alpha},
\end{align*}
which yields \eqref{eq:Gsing3d}.

The derivation of the 2d singular behavior in \eqref{eq:Gsing2d} is similar. In particular, suppose a function $u(x)$ satisfies \eqref{eq:pdeu} with the singularity 
\begin{align}\label{eq:usingderivation2d}
    u(x)
    \sim B\ln|x-x_0|\quad\text{as }x\to x_0\in\Omega\subset\R^2.
\end{align}
Integrating \eqref{eq:pdeu} over $\Omega_\eps$ and using the divergence theorem and the singularity in \eqref{eq:usingderivation2d} gives
\begin{align*}
    A\sim\int_{\Omega_\eps}f(x)\,\dd x+A
    &=\int_{\partial\Omega_\eps}D(x)^\alpha\nabla u\cdot \mathbf{n}\,\dd S\\
    &\sim 2\pi D(x_0)^\alpha B\quad\text{as }\eps\to0,
\end{align*}
and thus
\begin{align*}
    B
    =\frac{A}{2\pi D(x_0)^\alpha},
\end{align*}
which yields \eqref{eq:Gsing2d}.

\bibliography{library.bib}

\begin{thebibliography}{10}

\bibitem{grebenkov2024target}
Denis Grebenkov, Ralf Metzler, and Gleb Oshanin.
\newblock {\em Target search problems}.
\newblock Springer, 2024.

\bibitem{shoup82}
David Shoup and Attila Szabo.
\newblock Role of diffusion in ligand binding to macromolecules and cell-bound
  receptors.
\newblock {\em Biophys J}, 40(1):33, 1982.

\bibitem{meerson2015}
B~Meerson and S~Redner.
\newblock Mortality, redundancy, and diversity in stochastic search.
\newblock {\em Phys Rev Lett}, 114(19):198101, 2015.

\bibitem{kurella2015}
V~Kurella, J~C Tzou, D~Coombs, and M~J Ward.
\newblock Asymptotic analysis of first passage time problems inspired by
  ecology.
\newblock {\em Bull Math Biol}, 77(1):83--125, 2015.

\bibitem{benichou2008narrow}
O~B{\'e}nichou and R~Voituriez.
\newblock Narrow-escape time problem: Time needed for a particle to exit a
  confining domain through a small window.
\newblock {\em Physical review letters}, 100(16):168105, 2008.

\bibitem{grebenkov2016}
Denis~S Grebenkov.
\newblock Universal formula for the mean first passage time in planar domains.
\newblock {\em Phys Rev Lett}, 117(26):260201, 2016.

\bibitem{agranov2018narrow}
Tal Agranov and Baruch Meerson.
\newblock Narrow escape of interacting diffusing particles.
\newblock {\em Physical Review Letters}, 120(12):120601, 2018.

\bibitem{holcman2014}
D~Holcman and Z~Schuss.
\newblock The narrow escape problem.
\newblock {\em {SIAM} Rev}, 56(2):213--257, 2014.

\bibitem{ward10}
S.~Pillay, M.~J. Ward, A.~Peirce, and T.~Kolokolnikov.
\newblock {An asymptotic analysis of the mean first passage time for narrow
  escape problems: Part I: Two-dimensional domains}.
\newblock {\em Multiscale Model Simul.}, 8(3):803--835, 2010.

\bibitem{ward10b}
A.~F. Cheviakov, M.~J. Ward, and R.~Straube.
\newblock An asymptotic analysis of the mean first passage time for narrow
  escape problems: {P}art {II:} {T}he sphere.
\newblock {\em Multiscale Model Simul.}, 8(3):836--870, 2010.

\bibitem{PB3}
P~C Bressloff and S~D Lawley.
\newblock Escape from subcellular domains with randomly switching boundaries.
\newblock {\em Multiscale Model Sim}, 13(4):1420--1445, 2015.

\bibitem{lawley2019dtmfpt}
S~D Lawley and C~E Miles.
\newblock Diffusive search for diffusing targets with fluctuating diffusivity
  and gating.
\newblock {\em Journal of Nonlinear Science}, 2019.
\newblock https://doi.org/10.1007/s00332-019-09564-1.

\bibitem{bressloff2020search}
Paul~C Bressloff.
\newblock Search processes with stochastic resetting and multiple targets.
\newblock {\em Physical Review E}, 102(2):022115, 2020.

\bibitem{cherstvy2013anomalous}
Andrey~G Cherstvy, Aleksei~V Chechkin, and Ralf Metzler.
\newblock Anomalous diffusion and ergodicity breaking in heterogeneous
  diffusion processes.
\newblock {\em New Journal of Physics}, 15(8):083039, 2013.

\bibitem{cherstvy2014particle}
Andrey~G Cherstvy, Aleksei~V Chechkin, and Ralf Metzler.
\newblock Particle invasion, survival, and non-ergodicity in 2d diffusion
  processes with space-dependent diffusivity.
\newblock {\em Soft Matter}, 10(10):1591--1601, 2014.

\bibitem{massignan2014nonergodic}
Pietro Massignan, Carlo Manzo, Juan~A Torreno-Pina, Maria~F Garc{\'\i}a-Parajo,
  Maciej Lewenstein, and Gerald~J Lapeyre~Jr.
\newblock Nonergodic subdiffusion from brownian motion in an inhomogeneous
  medium.
\newblock {\em Physical review letters}, 112(15):150603, 2014.

\bibitem{xu2020heterogeneous}
Yong Xu, Xuemei Liu, Yongge Li, and Ralf Metzler.
\newblock Heterogeneous diffusion processes and nonergodicity with gaussian
  colored noise in layered diffusivity landscapes.
\newblock {\em Physical Review E}, 102(6):062106, 2020.

\bibitem{wang2020anomalous}
Wei Wang, Andrey~G Cherstvy, Xianbin Liu, and Ralf Metzler.
\newblock Anomalous diffusion and nonergodicity for heterogeneous diffusion
  processes with fractional gaussian noise.
\newblock {\em Physical Review E}, 102(1):012146, 2020.

\bibitem{wang2021time}
Wei Wang, Andrey~G Cherstvy, Holger Kantz, Ralf Metzler, and Igor~M Sokolov.
\newblock Time averaging and emerging nonergodicity upon resetting of
  fractional brownian motion and heterogeneous diffusion processes.
\newblock {\em Physical Review E}, 104(2):024105, 2021.

\bibitem{godec2015optimization}
Alja{\v{z}} Godec and Ralf Metzler.
\newblock Optimization and universality of brownian search in a basic model of
  quenched heterogeneous media.
\newblock {\em Physical Review E}, 91(5):052134, 2015.

\bibitem{vaccario2015}
G~Vaccario, C~Antoine, and J~Talbot.
\newblock First-passage times in $d$-dimensional heterogeneous media.
\newblock {\em Phys Rev Lett}, 115(24):240601, 2015.

\bibitem{godec2016first}
Alja{\v{z}} Godec and Ralf Metzler.
\newblock First passage time distribution in heterogeneity controlled kinetics:
  going beyond the mean first passage time.
\newblock {\em Scientific reports}, 6(1):20349, 2016.

\bibitem{english2011single}
Brian~P English, Vasili Hauryliuk, Arash Sanamrad, Stoyan Tankov, Nynke~H
  Dekker, and Johan Elf.
\newblock Single-molecule investigations of the stringent response machinery in
  living bacterial cells.
\newblock {\em Proceedings of the National Academy of Sciences},
  108(31):E365--E373, 2011.

\bibitem{kuhn2011protein}
Thomas K{\"u}hn, Teemu~O Ihalainen, Jari Hyv{\"a}luoma, Nicolas Dross, Sami~F
  Willman, J{\"o}rg Langowski, Maija Vihinen-Ranta, and Jussi Timonen.
\newblock Protein diffusion in mammalian cell cytoplasm.
\newblock {\em PloS one}, 6(8):e22962, 2011.

\bibitem{cutler2013multi}
Patrick~J Cutler, Michael~D Malik, Sheng Liu, Jason~M Byars, Diane~S Lidke, and
  Keith~A Lidke.
\newblock Multi-color quantum dot tracking using a high-speed hyperspectral
  line-scanning microscope.
\newblock {\em PloS one}, 8(5):e64320, 2013.

\bibitem{manzo2015review}
Carlo Manzo and Maria~F Garcia-Parajo.
\newblock A review of progress in single particle tracking: from methods to
  biophysical insights.
\newblock {\em Reports on progress in physics}, 78(12):124601, 2015.

\bibitem{wu2018}
Y~Wu, B~Han, Y~Li, E~Munro, D~J Odde, and E~E Griffin.
\newblock Rapid diffusion-state switching underlies stable cytoplasmic
  gradients in the caenorhabditis elegans zygote.
\newblock {\em Proc Natl Acad Sci}, page 201722162, 2018.

\bibitem{pacheco2024langevin}
Adrian Pacheco-Pozo, Micha{\l} Balcerek, Agnieszka Wy{\l}omanska, Krzysztof
  Burnecki, Igor~M Sokolov, and Diego Krapf.
\newblock Langevin equation in heterogeneous landscapes: how to choose the
  interpretation.
\newblock {\em Physical Review Letters}, 133(6):067102, 2024.

\bibitem{klimontovich1994nonlinear}
Yu~L Klimontovich.
\newblock Nonlinear brownian motion.
\newblock {\em Physics-Uspekhi}, 37(8):737, 1994.

\bibitem{bressloff2017temporal}
P~C Bressloff and S~D Lawley.
\newblock Temporal disorder as a mechanism for spatially heterogeneous
  diffusion.
\newblock {\em Phys Rev E - Rapid Comm}, 95(6):060101, 2017.

\bibitem{oksendal2003}
B.~Oksendal.
\newblock {\em Stochastic {Differential} {Equations}: {An} {Introduction} with
  {Applications}}.
\newblock Springer, 2003.

\bibitem{ito1944stochastic}
Kiyosi It{\^o}.
\newblock Stochastic integral.
\newblock {\em Proceedings of the Imperial Academy}, 20(8):519--524, 1944.

\bibitem{stratonovich1966new}
RL~Stratonovich.
\newblock A new representation for stochastic integrals and equations.
\newblock {\em SIAM Journal on Control}, 4(2):362--371, 1966.

\bibitem{hanggi1982stochastic}
Peter H{\"a}nggi and Harry Thomas.
\newblock Stochastic processes: Time evolution, symmetries and linear response.
\newblock {\em Physics Reports}, 88(4):207--319, 1982.

\bibitem{van1981ito}
Nicolaas~G Van~Kampen.
\newblock It{\^o} versus stratonovich.
\newblock {\em Journal of Statistical Physics}, 24(1):175--187, 1981.

\bibitem{volpe2010influence}
Giovanni Volpe, Laurent Helden, Thomas Brettschneider, Jan Wehr, and Clemens
  Bechinger.
\newblock Influence of noise on force measurements.
\newblock {\em Physical review letters}, 104(17):170602, 2010.

\bibitem{mannella2011comment}
Riccardo Mannella and PVE McClintock.
\newblock Comment on ``influence of noise on force measurements''.
\newblock {\em Physical review letters}, 107(7):078901, 2011.

\bibitem{volpe2011volpe}
Giovanni Volpe, Laurent Helden, Thomas Brettschneider, Jan Wehr, and Clemens
  Bechinger.
\newblock Volpe et al. reply.
\newblock {\em Physical Review Letters}, 107(7):078902, 2011.

\bibitem{bhattacharyay2025brownian}
A~Bhattacharyay.
\newblock Brownian motion near a wall: the dilemma of it{\^o} or stratonovich.
\newblock {\em Journal of Physics A: Mathematical and Theoretical},
  58(21):213001, 2025.

\bibitem{mannella2012ito}
Riccardo Mannella and Peter~VE McClintock.
\newblock It{\^o} versus stratonovich: 30 years later.
\newblock {\em Fluctuation and Noise Letters}, 11(01):1240010, 2012.

\bibitem{sokolov2010ito}
Igor~M Sokolov.
\newblock {Ito, Stratonovich, H{\"a}nggi and all the rest: The thermodynamics
  of interpretation}.
\newblock {\em Chemical Physics}, 375(2-3):359--363, 2010.

\bibitem{redner2001}
Sidney Redner.
\newblock {\em A guide to first-passage processes}.
\newblock Cambridge University Press, 2001.

\bibitem{tung2025escape}
Hwai-Ray Tung and Sean~D Lawley.
\newblock Escape from heterogeneous diffusion.
\newblock {\em arXiv preprint arXiv:2512.19646}, 2025.

\bibitem{ward1993summing}
M.~J. Ward, W.~D. Heshaw, and J.~B. Keller.
\newblock Summing logarithmic expansions for singularly perturbed eigenvalue
  problems.
\newblock {\em SIAM J. Appl. Math}, 53(3):799--828, 1993.

\bibitem{ward1993strong}
M~J Ward and J~B Keller.
\newblock Strong localized perturbations of eigenvalue problems.
\newblock {\em SIAM J Appl Math}, 53(3):770--798, 1993.

\bibitem{grebenkov2006}
D~S Grebenkov.
\newblock Partially reflected brownian motion: a stochastic approach to
  transport phenomena.
\newblock {\em Focus on probability theory}, pages 135--169, 2006.

\bibitem{cheviakov2010asymptotic}
Alexei~F Cheviakov, Michael~J Ward, and Ronny Straube.
\newblock {An asymptotic analysis of the mean first passage time for narrow
  escape problems: Part II: The sphere}.
\newblock {\em Multiscale Modeling \& Simulation}, 8(3):836--870, 2010.

\bibitem{cheviakov2011optimizing}
Alexei~F Cheviakov and Michael~J Ward.
\newblock Optimizing the principal eigenvalue of the laplacian in a sphere with
  interior traps.
\newblock {\em Mathematical and Computer Modelling}, 53(7-8):1394--1409, 2011.

\bibitem{jackson1975}
J~D Jackson.
\newblock {\em Classical {Electrodynamics}}.
\newblock Wiley, New York, 2nd edition edition, October 1975.

\bibitem{kolokolnikov2005}
Theodore Kolokolnikov, Michele~S Titcombe, and Michael~J Ward.
\newblock Optimizing the fundamental neumann eigenvalue for the laplacian in a
  domain with small traps.
\newblock {\em European Journal of Applied Mathematics}, 16(2):161--200, 2005.

\bibitem{erban2007reactive}
R.~Erban and S.~J. Chapman.
\newblock Reactive boundary conditions for stochastic simulations of
  reaction-diffusion processes.
\newblock {\em Phys Biol}, 4(1), 2007.

\end{thebibliography}
\bibliographystyle{unsrt}

\end{document}